\documentclass{pasj00}
\begin{document}
\SetRunningHead{Makino {\it et al.}}{GRAPE-6}
\Received{2000/12/31}%{yyyy/mm/dd}
\Accepted{2001/01/01}%{yyyy/mm/dd}

\title{GRAPE-6: The massively-parallel special-purpose computer for
astrophysical particle simulations}

%%% begin:list of authors
\author{Junichiro \textsc{Makino}}%
\affil{Department of Astronomy, School of Science, University of Tokyo, 
 Tokyo 133-0033, Japan}
\email{makino@astron.s.u-tokyo.ac.jpa}

\author{Toshiyuki  \textsc{Fukushige},
Masaki  \textsc{Koga}}
\affil{Department of
General System Studies, College of Arts and Sciences,
University of Tokyo,\\ Tokyo 153-8902, Japan}
\email{fukushig@provence.c.u-tokyo.ac.jp}
\and
\author{Ken {\sc Namura}}
\affil{IBM Japan Industrial Solution Co.,Ltd}\email{JL26165@jp.ibm.com}
%%% end:list of authors

%%% Please use the following style in case that sorting by 
%%% affilation is impossible. 
%
% \author{%
%   D-Firstname \textsc{D-Familyname}\altaffilmark{1}
%   E-Firstname \textsc{E-Familyname}\altaffilmark{1,2}
%   and
%   F-Firstname \textsc{F-Familyname}\altaffilmark{2}}
% \altaffiltext{1}{Address of Institute}
% \email{ddddd@xxx.xxx.xx.xx}
% \email{eeeee@xxx.xxx.xx.xx}
% \altaffiltext{2}{Address of Institute}

%% `\KeyWords{}' always has to be placed before `\maketitle'.
 %Do NOT move this preamble from here!

\maketitle

\begin{abstract}

In this paper, we describe the architecture and performance of the
GRAPE-6 system, a massively-parallel special-purpose computer for
astrophysical $N$-body simulations. GRAPE-6 is the successor of
GRAPE-4, which was completed in 1995 and achieved the theoretical peak
speed of 1.08 Tflops. As was the case with GRAPE-4, the primary
application of GRAPE-6 is simulation of collisional systems, though it
can be used for collisionless systems. The main differences between
GRAPE-4 and GRAPE-6 are (a) The processor chip of GRAPE-6 integrates 6
force-calculation pipelines, compared to one pipeline of GRAPE-4
(which needed 3 clock cycles to calculate one interaction), (b) the
clock speed is increased from 32 to 90 MHz, and (c) the total number
of processor chips is increased from 1728 to 2048. These improvements
resulted in the peak speed of 64 Tflops. We also discuss the design of
the successor of GRAPE-6.

\end{abstract}

%\newpage
%\tableofcontents
%\newpage

\section{Introduction}

The $N$-body simulation technique, in which the equations of motion
of $N$ particles are integrated numerically, has been one of the most
powerful tools for the study of astronomical objects such 
as the solar system, star clusters, galaxies, clusters of galaxies and 
large-scale structures of the universe.

Roughly speaking, the target systems for $N$-body simulations can be
classified into two categories: collisional systems and collisionless
systems. In the case of collisional systems, the evolution of the
system is driven by two-body relaxation process, in other words, by
microscopic exchange of thermal energies between particles. In this
case, the simulation timescale tends to be long, since the relaxation
timescale measured by the dynamical timescale is proportional to
$N/\log N$, where $N$ is the number of particles in the system.

The calculation cost of the simulation of collisional systems
increases rapidly as we increase the number of particles $N$, because
of the following two reasons. First, as stated above, the relaxation
timescale increases roughly linearly as we increase $N$. This means
the number of timesteps also increases at least
linearly\citep{MakinoHut1988}. The second reason is that it is not
easy to use fast and approximate algorithms such as Barnes-Hut tree
algorithm\citep{BarnesHut1986} or the fast multipole
method\citep{GreengardRokhlin1987} to calculate the interaction
between particles. Those imply that the cost per timestep is $O(N^2)$,
and that the total cost of the simulation is $O(N^3)$.

There are two reasons why the use of approximate algorithms for the force
calculation is difficult. The first reason is the need
for  relatively high accuracy. Since the total number of timesteps is
very large, we need a rather high accuracy for the force
calculation. The other reason is the wide difference in the orbital
timescale of particles. A  unique nature of the gravitational
$N$-body problem is that particles interact only through gravity,
which is an attractive force. This means that two particles can
approach arbitrary close during a hyperbolic close encounter. In
addition, spatial inhomogeneity tends to develop, resulting in  a high-density
core and a low-density halo. Even on average, particles in
the core require much smaller timesteps than particles in the halo do.

It is clearly very wasteful to apply the same timestep to all
particles in the system, and it is crucial to be able to apply
individual and adaptive timestep to each particle. Such an
``individual timestep'' algorithm, first developed by Aarseth
(\yearcite{Aarseth1963,Aarseth1999b}), has been the core for practically any
program that handles the time integration of collisional $N$-body
systems such as star clusters and systems of planetesimals.

The basic idea of the individual timestep algorithm is to assign
different times and timesteps to particles in the system. For particle
$i$, its next time is $t_i + \Delta t_i$, where $t_i$ is the current
time and $\Delta t_i$ is the current timestep. To integrate the
system, we first chose a particle with minimum $t_i+\Delta t_i$ and
set the current system time $t$ to be $t_i+\Delta t_i$. Then, we
predict the positions of all particles at time $t$ and calculate the
force on particle $i$. Finally, we correct the position of particle
$i$ using the calculated force, update $t_i$ and determine the new
timestep $\Delta t_i$. In practice, we force the size of timesteps to
be powers of two, so that the system time is quantized and multiple
particles have exactly the same time. In this way, we can use parallel
or vector processors efficiently, since we can integrate multiple
particles in parallel \citep{McMillan1986,Makino1991a}.

It is necessary to use the linear multistep method
(predictor-corrector method) with variable stepsize for the time
integration. Aarseth adopted an algorithm with third-order Newton
interpolation. Recently, the method based on the third-order Hermite
interpolation\citep{Makino1991d2,MakinoAarseth1992} has become widely
used, because of its simplicity.

In principle, it is not impossible to combine individual timestep
algorithm and fast algorithms such as Barnes-Hut tree algorithm or
FMM. McMillan and Aarseth (\yearcite{McMillanAarseth1993}) developed
such a combination, where the tree structure is dynamically updated
according to the move of particles and force is calculated using
multipole expansion up to octupole. They assigned predictor
polynomials to each node of the tree structure so that they could
calculate the force from nodes to particles at arbitrary times.

A serious problem with such a combination is that there is no known
method to implement it on parallel computers with distributed
memory. It is not simple to achieve a good parallel
performance with individual timestep algorithm, even without the tree
algorithm. The reason is that simple methods require fast and low-latency 
communication between processors. The recently proposed two-dimensional
algorithm \citep{Makino2002} somewhat relaxes the requirement for the
communication bandwidth, but it still requires low-latency
communication. When combined with the tree algorithm, efficient
parallelization becomes even more difficult.

Distributed-memory parallel computers have been used to run
large-scale cosmological simulations, with or without individual
timestep algorithm\citep{Dubinski1996,Springeletal2000}. In this
case, we use simple spatial decomposition to distribute particles over
processors. This works fine with large-scale cosmological simulations,
where the distribution of particles in large scale is almost
uniform. Many structures form from initial density fluctuations, and
many small high-density regions develop. Even so, we can still divide
the entire system so that the calculation load is reasonably well
balanced. In addition, the range of the timesteps is relatively small.

To parallelize the simulation of a single star cluster is much more
difficult, because the calculation cost is dominated by a small number
of particles in a single, small core\citep{MakinoHut1988}.
Therefore, communication latency becomes the bottleneck, and it is
difficult to parallelize the simple direct summation algorithm. As a
result, no good parallel implementation of the combination of the tree
algorithm and individual timestep algorithm exists. To really
accelerate calculation of a single cluster, we need an approach
different from what has been tried.

There are three different approaches to improve the speed of any
simulation: a) to use a faster computer, b) to use algorithms with
smaller calculation cost, and c) to improve the efficiency of the
algorithm used. Usually, option (a) means to use commercially
available fast computers, which, at present, means distributed-memory
parallel computers. An alternative possibility is to develop a
computer by ourselves. We have been pursuing this direction, starting
with GRAPE-1 \citep{Ito1990}.

The basic idea of the GRAPE (GRAvity piPE)
architecture \citep{Sugimotoetal1990}  is to develop
a fully pipelined processor specialized for the calculation of the
gravitational interaction between particles. In this way, a single
force-calculation pipeline integrates more than 30 arithmetic units,
which all operate in parallel. In the cause of the Hermite time
integration, we also need to calculate the first time derivative of the
force, resulting in nearly 60 arithmetic operations. This means that
we can integrate a large number of arithmetic unit into a single
hardware with minimal amount of additional logic. 

GRAPE-1 was an experimental hardware with very short word format
(relative force accuracy of 5\% or so), and not really suited for
simulations of collisional systems.  However, its exceptionally good
cost-performance ratio made it useful for simulations of collisionless
systems \citep{Okumura1991,Funato1992a}. Also, we developed an
algorithm to accelerate the Barnes-Hut tree algorithm using GRAPE
hardware \citep{Makino1991c}, and developed  GRAPE-1A
\citep{Fukushige1991}, which was designed to achieve good performance
with treecode. Thus, GRAPE approach turned out to be quite effective,
not only for collisional simulations but also for collisionless
simulations, and also for SPH simulations
\citep{Umemura1993,Steinmetz1996}. GRAPE-1A and its successors,
GRAPE-3 \citep{Okumura1993} and GRAPE-5 \citep{Kawaietal2000} have
been used by researchers worldwide for many different problems.

In this paper, we discuss GRAPE-6, our newest machine for the
simulation of collisional systems. We briefly summarize the history of
hardwares here.

GRAPE-2 \citep{Ito1991} adopted usual 64- and 32-bit floating point 
number format, and could be used with Aarseth's NBODY3 program.
After GRAPE-2, we developed GRAPE-3\citep{Okumura1993}, which is essentially an LSI
implementation of GRAPE-1. In GRAPE-1, arithmetic operations were
realized by fixed-point ALU chips and ROM chips, and in GRAPE-2 by
floating-point ALU chips. Thus, we needed several tens of LSIs to
realize a single pipeline. With GRAPE-3, we implemented a single
pipeline to a single custom LSI chip, and developed a board with 24
chips. In this way, we achieved the speed of 9 Gflops per board
(24 chips each performing 38 operations on 10 MHz clock cycle).

GRAPE-4\citep{Makinoetal1997} is similarly a single-LSI implementation
of GRAPE-2, or actually that of HARP-1\citep{Makinoetal1993}, which
was designed to calculate force and its time derivative. A single
GRAPE-4 chip calculates one interaction in every three clock cycles,
performing 19 operations. Its clock frequency was 32 MHz and peak
speed of a chip was 608 Mflops.

A major difference between GRAPE-4 and previous machines is its
size. GRAPE-4 integrated 1728 pipeline chips, for the peak speed of
1.08 Tflops. The machine is composed of 4 clusters, each with 9
processor boards. A single processor board houses 48 processor chips,
all of which share a single memory unit through another custom chip to 
handle predictor polynomials. GRAPE-4 chip uses two-way virtual
multiple pipeline, so that one chip looks like two chips with half the
clock speed. Thus, one GRAPE-4 board calculates the forces on 96
processors in parallel. Different boards calculate the forces from
different particles, but to the same 96 particles. Forces calculated
in a single cluster are summed up by special
hardware within the cluster.

In this paper, we describe the architecture and performance of
GRAPE-6, which is the direct successor of GRAPE-4. The main difference 
between GRAPE-4 and GRAPE-6 is in the performance. The GRAPE-6 chip
integrates 6 pipelines operating at 90 MHz, offering the speed of 30.8 
Gflops, and the entire GRAPE-6 system with 2048 chips offers the speed 
of 63.04 Tflops. 

The plan of this paper is as follows.  In section 2, we describe the
overall architecture, and in sections 3 and 4 the details of
implementation. In section 5, we discuss the difference between
GRAPE-4 and GRAPE-6. In section 6 we discuss the performance. Section
7 is for discussions. Those who are interested in how to use GRAPE-6,
but not much in the design details, could skip section 2.1, most of
section 3 and section 5.

\section{The architecture of GRAPE-6}

\def\br{\mathbf{r}}
\def\bx{\mathbf{x}}
\def\bv{\mathbf{v}}
\def\ba{\mathbf{a}}
\def\badot{\mathbf{ \dot a}}
\def\batwo{{\mathbf{a}^{(2)}}}
\def\adot{{ \dot a}}
\def\atwo{{{a}^{(2)}}}
\def\dt{{\Delta t}}
\def\bathree{{\mathbf{a}^{(3)}}}
\def\sub#1{_{\rm #1}}
\def\sup#1{^{\rm #1}}

In this section, we give the overview of the architecture of GRAPE-6.
What GRAPE-6 calculates are the following. First, it calculates the
gravitational force, its time derivative, and potential, given by
equations
\begin{eqnarray}
\label{eqn:force}
\ba_i =& \sum_j Gm_j\displaystyle{{\br}_{ij} \over (r_{ij}^2+\epsilon^2)^{3/2}},\\
\label{eqn:adot}
\badot_i =& \sum_jGm_j\left[
\displaystyle{{\bv}_{ij} \over (r_{ij}^2+\epsilon^2)^{3/2}} -
{3({\bv}_{ij}\cdot {\br}_{ij}) {\br}_{ij} \over
(r_{ij}^2+\epsilon^2)^{5/2}}\right],\\
\label{eqn:phi}
\phi_i =& \sum_j Gm_j\displaystyle{1 \over (r_{ij}^2+\epsilon^2)^{1/2}},
\end{eqnarray}
where  $\ba_i$, $\badot_i$, and $\phi_i$ are
the gravitational acceleration,  its first time derivative, and the potential  of particle
$i$, $m_i$, $\bx_i$ and $\bv_i$ are the mass, position and velocity of
particle $i$, $G$ is the gravitational constant and $\epsilon$ is the
softening parameter. GRAPE-6 hardware assumes $G=1$. If necessary, the
host computer can multiply the result calculated by GRAPE-6 by some
constant to use $G$ other than one. Also note that potential is calculated
without minus sign.   Relative position $\br_{ij}$ and relative velocity 
 and $\bv_{ij}$ are defined as
\begin{eqnarray}
   {\br}_{ij} &= {\bx}_j - {\bx}_i,\\
   {\bv}_{ij} &= {\bv}_j - {\bv}_i.
\end{eqnarray}
While calculating the force, it also evaluates the distance to the nearest neighbor
\begin{equation}
r_{min} = \min_{j\ne i} r_{ij},
\end{equation} 
and the value of index $j$ which gives the minimum distance. In addition,
it constructs the list of neighbor particles, whose distance squared
(with softening, $r_{ij}^2+\epsilon^2$)
is smaller than pre-specified value $h_{i}^2$.

The position $\bx_j$ and velocity $\bv_j$ of particles that exert the
forces are ``predicted'' by the following predictor polynomial

\begin{eqnarray}\label{eq:predictor}
    \bx_{j,{\rm p}} =& \displaystyle{\Delta t_j^4 \over 24}\batwo_{j,0} + 
\displaystyle{\Delta t_j^3 \over 6}\badot_{j,0} + 
    {\Delta t_j^2 \over 2}\ba_{j,0} + 
    \Delta t_j \bv_{j,0} + \bx_{j,0}\\
\label{eq:vpredictor}
\bv_{j,{\rm p}} =& \displaystyle{\Delta t_j^3 \over 6}\batwo_{j,0} + 
\displaystyle{\Delta t_j^2 \over 2}\badot_{j,0} + 
    \Delta t_j\ba_{j,0} + 
    \bv_{j,0},
\end{eqnarray}
where $\bx_{j,{\rm p}}$ and $\bv_{j,{\rm p}}$ are the predicted position and
velocity, $\bx_{j,0}$, $\bv_{j,0}$, $\ba_{j,0}$ and $\badot_{j,0}$ are the position,
velocity, acceleration and its time derivative of particle $j$ at time $t_{j,0}$,  and
$\Delta t_j$ is the difference between the current time $t_j$ of particle $j$
and system time $t$, $i. e., $
\begin{equation}
\Delta t_j = t - t_j.
\end{equation}

\subsection{Individual timestep on GRAPE hardware}

Here, we briefly summarize how GRAPE-6 (and GRAPE-4) works with
individual timestep algorithm. For more detailed discussion, see
\citet{Makinoetal1997} or \citet{MakinoTaiji1998}.

The time integration proceeds in the following
steps

\begin{description}
\item{a)} As the initialization procedure, the host sends all data (position,
velocity, acceleration, its first time derivative, mass and  time)
of all particles  to GRAPE memory unit.
\item{b)} The host creates the list of particles to be integrated at
the present timestep.
\item{c)} For each particles in the list, repeat the steps (d)-(g).
\item{d)} The host predicts the position and velocity of the particle,
and sends them to GRAPE. GRAPE stores them in the registers of the force 
calculation pipeline. It also sets the current time to a
register in the predictor pipeline.
\item{e)} GRAPE calculates the force from all other particles.  Positions and velocities of other particles at the current
time are calculated in the predictor pipeline.
\item{f)} After the calculation is finished, the host retrieves the
result. 
\item{g)} The host integrates the orbits of the particles and
determines new timesteps.
\item{h)} Update the present system time and go back to step (b).
\end{description}

Here, the key to achieve good performance is to send only particles
updated in the current timestep to GRAPE hardware. Thus, GRAPE
hardware need to have the memory unit large enough to keep all
particles in the system. This is usually not a severe limitation,
since even with fast GRAPE hardwares, the number of particles we can
handle with direct summation algorithm is not very large.

\subsection{Top-level network architecture}

\begin{figure}
\begin{center}
\leavevmode
\FigureFile(6 cm,6 cm){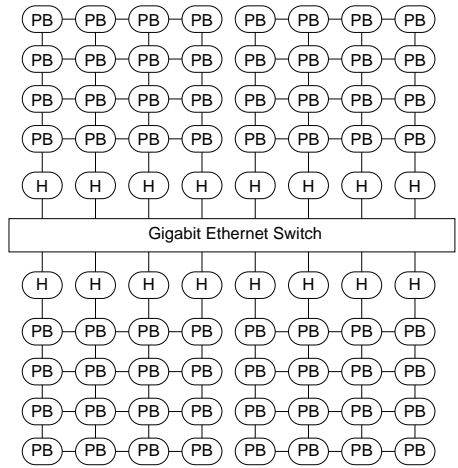}
\end{center}
\caption{The top level network structure of GRAPE-6. ``H'' indicates a
host computer and ``PB'' indicates a processor board.}
\label{fig:toplevel}
\end{figure}

The top-level architecture of GRAPE-6 is shown in figure
\ref{fig:toplevel}. It consists of 4 ``clusters'', each of which comprises 16
GRAPE-6 processor boards (PB), 4 host computers (H), and interconnection
networks. These 4 clusters are connected by Gigabit Ethernet.
For host computers, we currently use PCs with AMD
Athlon XP 1800+ CPU and SiS 745 chipset. Ethernet cards are 1000BT
cards with NS 83820 single-chip Ethernet controllers.

In the following, we will describe how we run parallel program on
GRAPE-6. First, let us concentrate on the parallelization within a
cluster.

\begin{figure}
\begin{center}
\leavevmode
\FigureFile(5 cm,5 cm){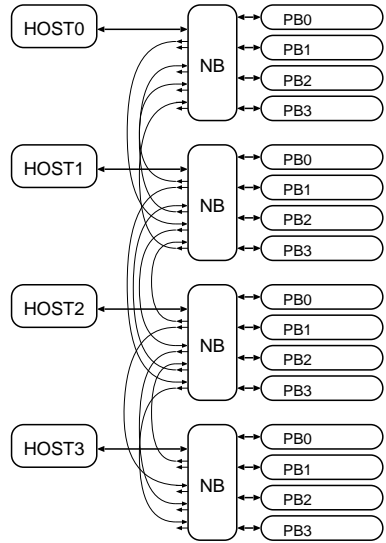}
\end{center}
\caption{A GRAPE-6 cluster. ``NB'' indicates a network board.}
\label{fig:g6cluster}
\end{figure}

Figure \ref{fig:g6cluster} shows one cluster.  Four processor boards
are connected to a host computer through a network board. Four network
boards are connected to each other, so that we can use a cluster as
single unit or as multiple units.

First consider the simplest case, where we just use 4 hosts to run
independent calculations. In this case,  4 processor boards
connected to a host through one network board calculate the forces on
the same set of particles, but from different set of particles [what
we called $j$-parallelism in \citet{Makinoetal1997}]. Each processor
board stores the different subset of particles in the particle memory,
and calculates the forces on the particles stored in the registers in
the processor chips.  The partial forces calculated in different
boards are sent in parallel to the network board, where they are added
together by an adder tree. The host computer receives the 
summed-up forces.  As will be discussed later, multiple processor
chips on one board also have their local memories to store
particles. They calculate the forces on the same set of particles, but
from different sets of particles. The partial forces are summed up by
the adder tree on the processor board. From the logical point of
view there is no difference between a single-board system and
multi-board system, as far as we use a single host. We can regard the
entire system just as a huge adder tree with processor chips at all leaves.

When all 16 boards and 4 hosts are used as a single unit, the
particles are divided to 4 groups and each group is assigned to one
host. Conceptually, the $j$-th board connected to host $i$ calculates
the force on particles in host $i$, from particles in host
$j$. Summation of the partial forces is performed in the same way as
in the case of single-host calculation. The only difference is
that the data to be stored in the memory come from other hosts.

\begin{figure}
\begin{center}
\leavevmode
\FigureFile(6 cm,6 cm){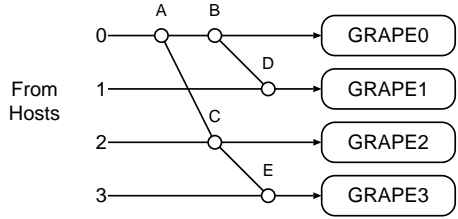}
\end{center}
\caption{A 4-input example of switching network for parallel GRAPE.}
\label{fig:scalablenet}
\end{figure}

In order to allow both single-host and multi-host calculations, the
network board must switch between broadcast mode (for the single-host
calculation) and point-to-point mode (for the multi-host
calculation). It would also be useful if we can use two hosts
together. In this case, it is necessary to accept two inputs, and to pass
each of them to two boards. Thus, we need three operation modes for
the network board. One simple way to implement these three modes is
shown in figure \ref{fig:scalablenet}. Here, nodes A and B simply
output the inputs from the left-hand side ports to two output ports. Nodes C,
D, and E can select one from two inputs. In the case of node C, the
selected input is sent to two output ports.

This network can be configured in three ways. In the first mode, all
nodes select input from the lower ports in figure
\ref{fig:scalablenet}. In other words, C takes input from input port
2, D from input port 1, and E from input port 3.  In this case, each
GRAPE receives data from the input port with the same index. In this
mode we can use this 4-GRAPE network as part of 4-host, 16-GRAPE
system.  In the second mode, node C selects the input from port 2,
while D and E selects data from upper input port in figure
\ref{fig:scalablenet} (nodes B and C for nodes D and E,
respectively). In this mode, GRAPEs 0 and 1 receive the same data from
port 0. Similarly, GRAPEs 2 and 3 receive the data from port 2. In
other words, GRAPEs 0 and 1 (and 2 and 3 as well) are effectively
bundled together to behave as one system, and we can use this system
as a part of 2-host, 8-GRAPE system. In the third mode, all nodes
select upper inputs, thereby sending the data from port 0 to all
GRAPEs. In this way, we can use this 4-GRAPE network as a single
system connected to one host.

An important character of this network is that its hardware cost is
$O(p)$, where $p$ is the number of GRAPE hardwares. Thus, even for
very large systems, the cost of the network remains small.
By using this hardware network to send data from multiple host to
processor boards under one host in parallel, we can improve the
parallel efficiency quite significantly.

There are many possible algorithms to parallelize the calculation over
multiple clusters. Here we show just one example, which is a
generalization of the ``copy'' algorithm \citep{Makino2002}. In the
copy algorithm, each node has the complete copy of the system. At each
timestep, each node, however, integrates its own share of particles,
which is either statically or dynamically assigned to it. After one
step is finished, all nodes broadcast particles they updated, so that
all nodes have the same updated system. In the case of multi-cluster
calculation, each cluster has a complete copy of the system, which is
distributed to 4 hosts. For example, host 0 of cluster 0 and
host 0 of cluster 1 have the same data. In the time integration,
calculation load is divided between all hosts in the different
clusters with the same internal index. After one step is finished,
updated data are exchanged again between hosts in different clusters
with the same index.  One could use ``ring'' algorithm or 2-D
algorithm \citep{Makino2002}, but for 4 clusters the difference in
the performance is rather small.

In principle, we could extend the network board to form 8-input,
8-output switch, so that we can use all 64 boards as a ``single
cluster''. We decided not to do this since for many scientific
applications we will use the system as a correction of single-host
systems to run multiple simulations independently. To run multiple
calculations, it is more efficient to have larger number of host
computers.

\subsection{board-level structure}

\begin{figure}
\begin{center}
\leavevmode
\FigureFile(\columnwidth,\columnwidth){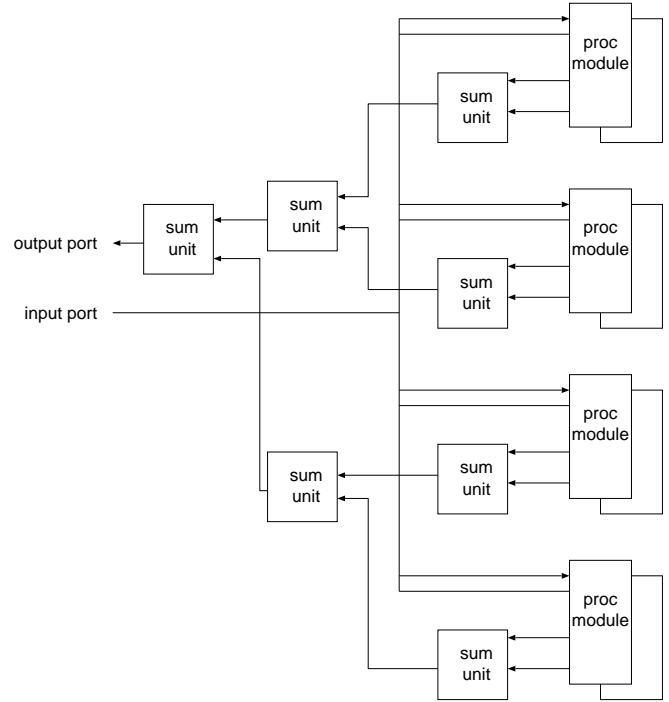}
\end{center}
\caption{The structure of the processor board.}
\label{fig:processorboard}
\end{figure}

Figure \ref{fig:processorboard} shows the structure of a processor
board. It houses 8 processor modules. The processor board has one
broadcast network that broadcasts data from the input port to all
processor modules, and one reduction network that reduces the results
obtained on 32 chips and returns it to the host through the output
port.

\begin{figure}
\begin{center}
\leavevmode
\FigureFile(6 cm,6 cm){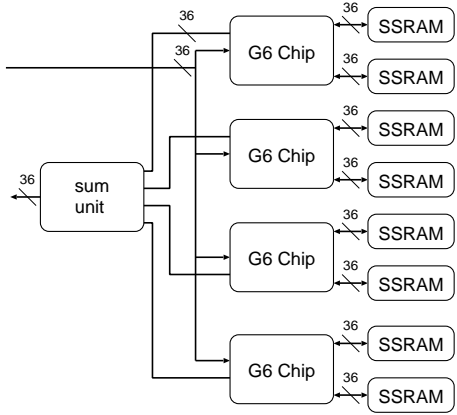}
\end{center}
\caption{The structure of the processor module.}
\label{fig:processormodule}
\end{figure}

Each processor module consists of 4 processor chips
each with its memory, and one summation unit. The structure of a
processor module is the same as that of the processor board, except
that it has 4 processor chips instead of 8 processor modules. Figure
\ref{fig:processormodule} shows the structure of a processor module. 

The memories attached to one processor chip can store up to 16,384 
particles. Thus, a single board with 32 chips can handle up to 524,288
particles, for direct summation code with individual timestep. A $4
\times 4$-board cluster can handle up to 2 million particles. If one
wants to use more than 2 million particles with direct summation, it
is possible to use the ring algorithm (see section 5.2). The calculation
with 8 million particles is theoretically possible on a single cluster
with 16 processor boards.

In the next two sections, we present the detailed description of the
hardware, in a bottom-up fashion. In section \ref{sect:chip}, we
describe the processor chip and in section \ref{sect:pbnh} the
processor board, network board and interconnection.

\section{The processor chip}
\label{sect:chip}

The GRAPE-6 processor chip was fabricated using Toshiba TC-240 process
(nominal design rule of $0.25 {\rm \mu m}$. The physical size of the
chip is roughly 10 mm by 10 mm, and packaged into 480-contact BGA
package. It operates at 90 MHz clock cycle. Power supply voltage is
2.5V. Heat dissipation is around 12 W at the maximum.

\begin{figure}
\begin{center}
\leavevmode
\FigureFile(7 cm,7 cm){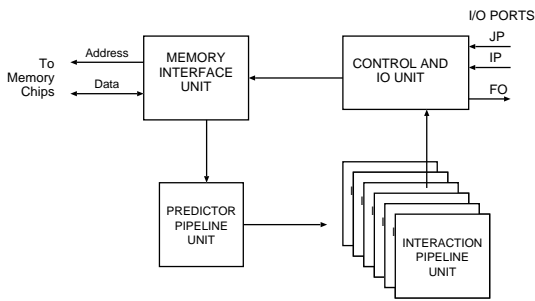}
\end{center}
\caption{The block diagram of the processor chip.}
\label{fig:processorchip}
\end{figure}

A processor chip consists of six force calculation pipelines, a
predictor pipeline, a memory interface, a control unit and I/O ports. Figure
\ref{fig:processorchip} shows the overview of the chip.  In the
following, we discuss each block in turn.

\subsection{Force calculation pipeline}

The task of the force calculation pipeline is to evaluate equations
(\ref{eqn:force})--(\ref{eqn:phi}). It also determines the nearest neighbor
particle and its distance. This function is rather convenient for
detecting close encounters or physical collisions between particles
that require special treatments. For this purpose, the indices of
particles that exert forces are supplied to the pipeline.

The indices are also used to avoid self-interaction. The force
calculation pipeline has the register for the index of the particle
for which the force is calculated, and avoids the accumulation of the
result if two indices are the same. This capability is introduced to
avoid the need to send particles twice to the memory in the case of
the individual timestep algorithms.

With the individual timestep algorithm and the hardwired predictor
pipeline, the data of particles which exert forces are evaluated by
the predictor pipeline on chip, while the data for the particle for
which the force is calculated is evaluated on the host computer and
sent to the register of the force calculation pipeline. These two
values are not exactly the same, since the data format and accuracy of
the hardware predictor are different from that of the host
computer. GRAPE-4 pipeline did not have the logic to use the particle
index, and the only way to avoid the self interaction was to make the
data exactly the same. To achieve this, for the particles to be
updated, we sent the predicted data at the current time to the memory
as well as the registers. This means that we had to send $j$-particles twice
per timestep. With the index-based approach, we need to send
$j$-particles only once per timestep, resulting in a significant
reduction in the total amount of communication.

For GRAPE-6 pipeline, we adopted the 8-way VMP [virtual multiple
pipeline, \citet{Makinoetal1997}], in which single physical pipeline
serves as eight virtual pipelines, calculating the forces on 8
different particles. In this way, we can reduce the requirement for
the memory bandwidth by a factor of 8, since all VMPs (and also
physical multiple pipelines on a chip) calculate the forces from the
same particle.

In the physical implementation of the pipeline, we adopted several
different number representations, depending on the required accuracy.
For input position data, we used 64-bit fixed point format. The reason
we used the fixed point format here is to simplify the
hardware. Additional advantage of using the fixed-point format is that
the implementation of the periodic boundary condition is simpler than
that in the case of the floating-point data format\citep{Fukushige1996}.

After first subtraction between two position vectors, the result is
converted to floating-point format with 24-bit mantissa. Here,
floating-point format is preferred, since otherwize we need very large
multipliers.

For the final accumulation, we return to the 64-bit fixed-point
format, again to simplify the hardware. Here, we specify the scaling
factor for each particle, so that we can calculate the forces with
very different magnitude, without causing overflow or underflow.

The pipeline for the calculation of the time derivative is designed in
a similar way, but with 20-bit mantissa for intermediate data and
32-bit fixed-point format for the final accumulation. Since the time
derivative is one order higher than the force, the required accuracy
is lower.

\begin{figure}
\begin{center}
\leavevmode
\FigureFile(\columnwidth,6 cm){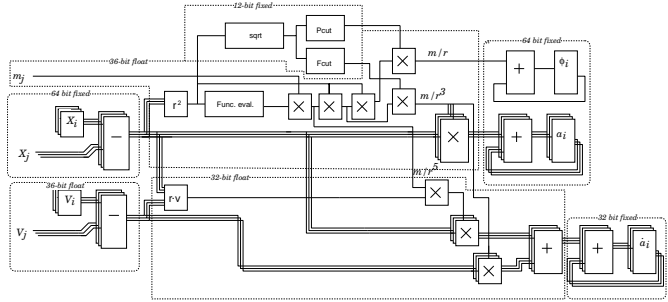}
\end{center}
\caption{The block diagram of the force calculation pipeline.}
\label{fig:forcepipe}
\end{figure}

Figure \ref{fig:forcepipe} shows the block diagram of the
pipeline. It consists of arithmetic units to perform the operations
shown in table \ref{tab:forcepipe}

\begin{table}
  \caption{Arithmetic operations in force calculation pipeline }\label{tab:forcepipe}
  \begin{center}
    \begin{tabular}{llccc}
\hline
\# & Operation & Format & length & mantissa \\
\hline
1& $ d\bx \leftarrow \bx_j  - \bx_i$ & fixed & 64 & --\\
2& $ dr_s^2  \leftarrow |d\bx|^2+\epsilon^2$ & float& 36 & 24\\
3& calculation of $ r_s^{-\alpha}$,  & float& 36 & 24\\
&  where $\alpha=1,3,5$ \\
4& $ \phi_{ij} \leftarrow m_jr_s^{-1}$ & float& 36 & 24\\
5& $ \phi_i \leftarrow \phi_i + \phi_{ij}$ & fixed & 64 & --\\
6& $ \ba_{ij} \leftarrow m_jr_s^{-3}d\bx$&float& 36 & 24\\
7& $ \ba_i \leftarrow \ba_i + \ba_{ij}$&  fixed & 64 & --\\
8& $ d\bv \leftarrow \bv_j  - \bv_i$&float& 36 & 24\\
9&  $ s\leftarrow d\bv \cdot d\bx$ &float& 32 & 20\\
10&  $ j_1\leftarrow d\bx \cdot 3s m_jr_s^{-5}$ &float& 32 & 20\\
11&  $ j_2\leftarrow d\bv \cdot m_jr_s^{-5}$ &float& 32 & 20\\
12&  $\badot_{ij} \leftarrow j_1 + j_2$ &float& 32 & 20\\
13&  $\badot_{i} \leftarrow \badot_i + \badot_{ij}$ & fixed & 32 & --\\
\hline
    \end{tabular}
  \end{center}
\end{table}

% \begin{enumerate}
% \item $ d\bx \leftarrow \bx_j  - \bx_i$
% \item $ dr_s^2  \leftarrow |d\bx|^2+\epsilon^2$
% \item calculation of $ r_s^{-\alpha}$, where $\alpha=1,3,5$
% \item $ \phi_{ij} \leftarrow m_jr_s^{-1}$
% \item $ \phi_i \leftarrow \phi_i + \phi_{ij}$
% \item $ \ba_{ij} \leftarrow m_jr_s^{-3}d\bx$
% \item $ \ba_i \leftarrow \ba_i + \ba_{ij}$
% \item $ d\bv \leftarrow \bv_j  - \bv_i$
% \item  $ s\leftarrow d\bv \cdot d\bx$
% \item  $ j_1\leftarrow d\bx \cdot 3*s* m_jr_s^{-5}$
% \item  $ j_2\leftarrow d\bv \cdot m_jr_s^{-5}$
% \item  $\badot_{ij} \leftarrow j_1 + j_2$
% \item  $\badot_{i} \leftarrow \badot_i + \badot_{ij}$
% \end{enumerate}

We briefly discuss each operations below.

\subsubsection{$ d\bx \leftarrow \bx_j  - \bx_i$}

The position data are expressed in the 64-bit 2's complement
fixed-point format.  The result of the subtraction is then converted
to the floating-point format. The floating-point format used here
consists of the sign bit, the zero bit, 10-bit exponent and 24-bit
mantissa. The sign bit expresses the sign (one denotes a
negative number). The zero bit indicates if the number is zero or not
(one indicates that the expressed value is zero). In
standard IEEE floating-point format, a zero value is expressed in a
special format (both exponent and mantissa are zero). This convention
is useful to achieve the maximum accuracy for a given word
length. However, using zero bit is more cost-effective in the internal
expression for the hardware, since the logic to handle zero value is
greatly simplified.

For the result of subtraction, the range of exponent is 6 bits. We use 
a biased format, and extend the exponent to 10 bits. For all
floating-point operations, we use 10-bit exponents, to avoid
overflows in intermediate results (in particular for $r^{-5}$).

The length of the mantissa is 24 bits, with usual ``hidden bit'' at
MSB (most significant bit). For the rounding mode, we used the ``force-1'' rounding, with the 
correction to achieve unbiased rounding. With the ``force-1''
rounding, we always set the LSB of the calculated result (after proper 
shifting) to be one, regardless of the contents of the field below
LSB. Thus, if the LSB is already one, the result is rounded toward
zero, and if the LSB is zero, the result is rounded toward
infinity. Thus, this rounding gives almost unbiased result.

However, in this simple form this rounding is still biased, since the
treatment for the case where all bits below LSB are zero is not
symmetric. Consider the following example, where we use 4 bits for
mantissa and calculated result is in 8 bits (for example with
multiplication). If the result is 10010000 in binary format, it is
rounded to 1001. If the result is 10000000, it is also
rounded to 1001. Thus, out of 32 possible combinations of
LSB bit and 4 bits under LSB, for 16 cases the rounding is upward, 
15 cases downward, and one case no change. This gives slight upward
bias for the rounded result.

One way to remove this bias is not to force one if all bits below LSB
is zero. This can be implemented with a rather simple logic, and we used 
this method with all floating-point arithmetic units used in GRAPE-6.

Compared to the usual ``round-to-the-nearest-even'' rounding, this
bias-corrected rounding is significantly easier to design and test. In
particular, there is no need for the conditional incrementer that
would be necessary with the usual nearest rounding. Of course, this
simpler design does not mean smaller number of gates, since our
rounding requires the length of the mantissa longer than that for the
nearest rounding by 1 bit. However, it is also true that the additional
number of gates is rather small, because we do not need the
conditional incrementer.

\subsubsection{ $ dr_s^2  \leftarrow |d\bx|^2+\epsilon^2$}

These are realized with usual floating-point multipliers and
adders. The design of the adder used here is simpler than
that of general-purpose floating-point adders, since we know that both
operands are positive. This means that the result of the addition of
the two mantissas is always larger than the larger of the two
operands, and we need to shift the result at the maximum by one
bit. With general-purpose adder, if two operands have similar
magnitudes and different signs, the result of the addition can be much
smaller, and we need a shifter with the capability to shift up the
result by up to the length of the mantissa itself.

\subsubsection{calculation of $ r_s^{-\alpha}$}

Here, we followed the design of the GRAPE-4 chip, where we used
segmented second-order polynomial to calculate  $ r_5= r_s^{-5}$. We then
multiply $r_5$ by $m_j$, and then by $r_s^2$ twice to obtain 
$m_jr_s^{-3}$  and $m_j/r_s$.

To calculate $r_s^{-5}$, we first normalize $r_s^2$ to the range of
$[1/4,1)$. In other words, $r_s$ is expressed as $2^{2a+b}\cdot c$,
where $a$ is an integer number, $b$ is either 0 or 1, and $c$ is in
the range of $[1/2,1)$. The ``exponent'' $a$ is multiplied by $-5$ to
obtain the exponent of the resulted $r_s^{-5}$.

We used a table with 512 entries to obtain the coefficients for the
polynomial. This table accepts $b$ and eight MSB bits (excluding the
hidden bit) as the input address. The output of the table consists of
the coefficients for the second-order term (12 bits), first-order term
(18 bits), zeroth-order term (24 bits) and exponent (3 bits). Note
that the calculated result is always smaller than the zeroth order
term, since both the first and second derivatives have minus
signs. Therefore, the MSB of the calculated result can turn to zero,
even though MSB of the zeroth order term is always one. In this case,
we need to shift the result by one bit, and adjust the exponent by
one.  This adjusted exponent is then added to previously calculated
exponent to obtain the exponent of the final result.

\subsubsection{ $ \phi_{ij} = m_jr_s^{-1}$}

As described in the previous subsection, we actually calculate $m_j
r_s^{-5}$ and then multiply it by $r_s^2$ twice. These multiplications are
usual floating-point multiplications, with the bias-corrected force-1
rounding.

\subsubsection{ $ \phi_i = \phi_i + \phi_{ij}$}

The potential is accumulated in the 64-bit fixed-point format. The
pairwise potential $\phi_{ij}$, which is obtained in the
floating-point format, is shifted before addition according to the
shift length $e_{\phi} - s_{\phi,i}$, where $e_{\phi}$ is the value of
exponent of the pairwise potential $\phi_{ij}$ and $s_{\phi,i}$ is the
scaling coefficient for the potential of particle $i$. Note that the
coefficient $s_{\phi,i}$ is specified on the per-particle basis and we
can specify different values for 48 virtual pipelines. This
coefficient should be calculated from a reasonably good estimate of
the total potential of particle $i$, to avoid both overflow and
underflow during calculation.

\subsubsection{ $ \ba_{ij} = m_jr_s^{-3}d\bx$}

These are   usual floating-point multiplications.

\subsubsection{ $ \ba_i = \ba_i + \ba_{ij}$}

Here, we use the same design as that for the accumulation of the
potential. The scaling coefficient is common for all three components.

\subsubsection{ $ d\bv \leftarrow \bv_j  - \bv_i$}

This and the remaining operations to be discussed in this section are
all for the time derivative of the forces. For these operations, we
use the number format with the 20-bit mantissa. For this first
subtraction, the mantissa  of input is 24 bits, and the result
is given with the 20-bit mantissa.

\subsubsection{  $ s= d\bv \cdot d\bx$}

This is an inner-product of two vectors in three
dimensions. The mantissa of $d\bx$ is first truncated to 20 bits. 

\subsubsection{  $\badot_{i} = \badot_i + \badot_{ij}$}

As in the case of the potential and the force, we use a fixed-point
format with scaling coefficient for the final accumulation. Instead of the
64-bit format, however, here we used  a 32-bit format, since only the
contributions from nearby particles are important for the time
derivative.

\subsubsection{Cutoff unit}

In figure \ref{fig:forcepipe}, there are three boxes in ``12-bit
fixed'' format region. These are used to implement the Gaussian cutoff
of the $1/r$ potential, to be used with Ewald summation method for the
calculation of the gravitational force with the periodic boundary
condition. The details of the  operation of these boxes will be
described elsewhere, with the discussion of the performance and
accuracy of the Ewald method on GRAPE-6. In this paper, we can regard
these two boxes, ``Pcut'' and ``Fcut'' as just boxes with constant
(unit) outputs.

\subsection{Neighbor list unit}

The neighbor list unit of GRAPE-6 chip is essentially the same as
that of GRAPE-4 board. It consist of two memory units, one for the
indices of $j$-particles and the other for flags to indicate (virtual) 
pipelines. One neighbor list unit serves 16 virtual pipelines (two
physical pipelines). Thus
we integrated three units to one chip. One neighbor list unit can
store up to 256 neighbor particles. 

\begin{figure}
\begin{center}
\leavevmode
\FigureFile(6 cm,6 cm){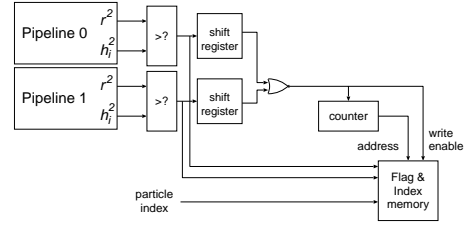}
\end{center}
\caption{The block diagram of the neighbor list unit.}
\label{fig:neighbor}
\end{figure}

Figure \ref{fig:neighbor} shows one neighbor list unit.  Each pipeline
has registers (for each of the virtual pipelines) for the neighbor
radius squared $h^2$, and if the distance to the current $j$-particle
is not larger than the neighbor radius, a flag is asserted. This flag
is stored to a shift register. Once per every eight clock cycles, this
shift register contains the eight flags from different VMPs for the
same $j$ particles. At this cycle, if any of 16 flags from 16 virtual
pipelines is asserted, the index of the current $j$-particle and the
flags themselves are written to the memory.

\subsection{Predictor pipeline}

The predictor pipeline evaluates the predictor polynomials expressed
in equations (\ref{eq:predictor}) and (\ref{eq:vpredictor}). As stated
earlier, we used 8-way VMP for the force calculation
pipeline. Therefore, the predictor pipeline can use eight clock cycles
to produce the predicted position and velocity of one particle. To
take advantage of this fact, we implemented one pipeline, which
processes $x$, $y$, and $z$ components sequentially. In principle, we
could further reduce the hardware size by using one pipeline for both
position and velocity. We did not adopt this approach since the
circuit size for the predictor pipeline was already a small fraction
of the total size of the chip.

In the design of the predictor pipeline, we tried to minimize the
amount of data to express the predictor for one particle, since it
directly affects the time needed for communication and the number of
wires needed between the memory chips and the processor chip.

With GRAPE-4, the  predictor data for one particle was expressed in 19 
32-bit words (2 for time, 6 for position, 3 for each of velocity,
acceleration, and time derivative of acceleration, 1 for mass, and one 
for memory address). With a similar format, GRAPE-6 predictor would
need 23 words, since we need one more word for particle index and
three more for the second time derivative of the acceleration. In many 
applications, inclusion of the second derivative improves the accuracy 
rather significantly.

To reduce the data length, we adopted the following two
methods. First, for the particle time, instead of sending the time
itself, we send the location of the bit below which the current system 
time $t$ can be  different from the particle time $t_j$. In this way, we could reduce 
the number of bits to express time from 64 to just 7.

Second, we use a block floating-point format with mantissa length
optimized for each of the predictor coefficients. Thus, we used 32,
20, 16 and 10 bits, for velocity, acceleration, first and second time
derivatives, respectively. 

With these two changes, we could make one predictor data to be
expressed in 16 32-bit words. Thus, we could use 64-bit memory bus
with clock speed the same as that for the pipeline, to supply one
particle data in every 8 clock periods.

\begin{figure}
\begin{center}
\leavevmode
\FigureFile(6 cm,6 cm){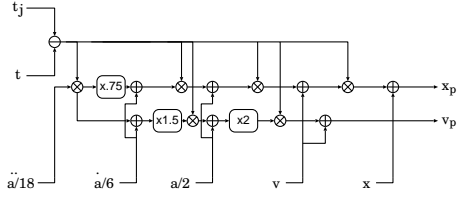}
\end{center}
\caption{The block diagram of the predictor unit.}
\label{fig:predictor}
\end{figure}

Figure \ref{fig:predictor} shows the block diagram of the predictor unit.

The predictor pipeline performs the following operations

\begin{description}
\item a) {$\Delta t \leftarrow t - t_j$}
\item b) {$ p_1 \leftarrow \Delta t \cdot (\atwo/18)$}
\item c) {$ p_2 \leftarrow p_1 \cdot 0.75$}
\item d) {$ p_3  \leftarrow p_2 + (\adot/6)$}
\item e) {$ p_4 \leftarrow p_3 \cdot \Delta t$}
\item f) {$ p_5  \leftarrow p_4 + (a/2)$}
\item g) {$ p_6 \leftarrow p_5 \cdot \Delta t$}
\item h) {$ p_7 \leftarrow p_6 + \bv$}
\item i) {$ p_8 \leftarrow p_7 \cdot \dt$}
\item j) {$ x_p  \leftarrow x + p_8$}
\item k) {$ q_1 \leftarrow p_1 + (\adot/6)$}
\item l) {$ q_2 \leftarrow q_1 \cdot 1.5$}
\item m) {$ q_3 \leftarrow q_2 \cdot \dt$}
\item n) {$ q_4 \leftarrow q_3 + (a/2)$}
\item o) {$ q_5 \leftarrow q_4 \cdot 2$}
\item p) {$ q_6 \leftarrow q_5 \cdot \dt$}
\item q) {$ v_p  \leftarrow v + q_6$}
\end{description}
Here, $p_i$ is the output of $i$-th arithmetic unit of the predictor
pipeline for the position, and $q_i$ is that for the
velocity. Notations like $(\atwo/18)$ mean the values corresponding
the expressions are supplied from the memory unit.

In the following, we describe operations a, b, c, d, j and q. Other
operations are simple fixed-point addition, multiplication, or
multiplication by a constant
implemented in the way similar to that for operations b, d and  c.

\subsubsection{$\Delta t \leftarrow t - t_j$}

\begin{figure}
\begin{center}
\leavevmode
\FigureFile(6 cm,6 cm){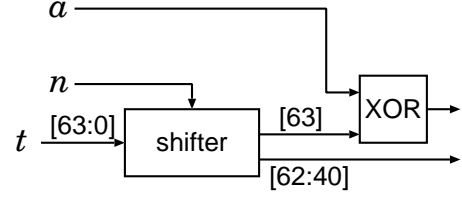}
\end{center}
\caption{The block diagram of the logic to handle subtraction of the time.}
\label{fig:dtcalc}
\end{figure}

The current system time $t$ is expressed in the 64-bit fixed-point
format. The particle time $t_j$ is expressed by the bit location $n$
above which $t$ and $t_j$ are the same. This location is the location
of  MSB of the timestep $\Delta t_j$. Consider the following
example. If $t_j= 0.5$ and $\Delta t_j= 0.125$, the current time $t$
must be in the range of $ [t_j, t_j+\Delta t_j]$, {\it i.e.}, $[0.5,
0.625]$. In this case, $t - t_j$ can be calculated by simply masking
all bits equal to or higher than MSB of $\Delta t_j$ ({\it i.e.},
0.125 and above). This works for any value of $t$ in the range
$[0.5,0.625)$.  However, if $t=0.625$, this procedure returns $0$,
but the correct value is $0.125$. This is simply because we masked the bit
which represents the exact value of $\Delta t_j$. This problem can be
solved by supplying the value of $t_j$ at that bit. Unless $t$ is
equal to $t_j + \Delta t_j$, values of this bit for $t$ and $t_j$ are
the same. In this case, the corresponding bit of  the resulting $\Delta t$ must be
0. However, if $t$ is equal to $t_j + \Delta t_j$, values of this bit
for $t$ and $t_j$ are different. In this case, MSB of the result must
be one. Thus, by taking XOR of the two input bits, we can determine
the MSB value of the result. Figure \ref{fig:dtcalc} shows the actual
circuit. Here, $a$ is the value of the bit of  $t_j$ which corresponds to
non-zero bit of $\Delta t_j$ and $n$ is the location of that bit. 
The result is expressed in a 24-bit unsigned fixed-point
format. Here, the rounding is simple rounding to zero. This can cause
very small bias in the predicted position, if the timestep of the
current blockstep is very small and the timestep of the predicted
particle is large. In this case, however, the error in the prediction
does not degrade the accuracy. Therefore we do not perform rounding
correction here.

\subsubsection{$ p_1 \leftarrow \Delta t \cdot (\atwo/18)$}

Both inputs are supplied in a 10-bit fixed-point format. Here, we use
the sign-magnitude format, instead of the usual 2's complement format, to
simplify the design of the multiplier. Note that $\Delta t$ is supplied in the 24-bit
format. Therefore we need to truncate it to 10-bit format, using the
bias-corrected force-1 rounding discussed earlier. The result is also
rounded to the 10-bit format.

\subsubsection{$ p_2 \leftarrow p_1 \cdot 0.75$}

This multiplication by constant is achieved by adding $p_1/2$ and
$p_1/4$. These two values can be calculated by shifting them to the
right by one bit and two bits, respectively. These shiftings, in hardware,
require just wiring and no logic. Thus, this multiplication is
actually implemented by a single adder.  Here we do not round the
result, since the effect of the error in the multiplication in the
predictor is usually very small.

\subsubsection{$ p_3  \leftarrow p_2 + (\adot/6)$}

Here, $p_2$ is in the 10-bit sign-and-magnitude format, and
$(\adot/6)$ is in a 16-bit format. Thus, we first extend $p_2$ to 16
bits. If two inputs have different signs, we need to determine which
one is larger. We do this by calculating both $a-b$ and $b-a$. If
$a-b$ does not cause overflow, we can see that $a\ge b$, and use $a-b$
as the result. The sign of the result is the same as that of $a$.
This circuit is rather complicated and larger than that for the 2's
complement format. However, the gate count is practically
negligible.

\subsubsection{Operations (e)-(i)}

All these are usual fixed-point addition or multiplication,
implemented in the same way as operations (b) and (d).

\subsubsection{$ x_p  \leftarrow x + p_8$}

Here, we add two numbers in different formats. One is $x$ in the
64-bit 2's complement fixed-point format. The other is $p_8$ in
floating-point format with sign, exponent and mantissa. Since we do
not perform any normalization during the calculation of $p_8$, the mantissa
is not normalized. This means that we do not use the hidden bit for
$p_8$. We first shift $p_8$ according to the value of the exponent of the
velocity and then add it to (or subtract it from) $x$ according to the
sign bit.

\subsubsection{Operations (k)-(p)}

These operations are implemented in the same way as similar operations
for the position predictor pipeline are implemented

\subsubsection {$ v_p  \leftarrow v + q_6$}

This is essentially the same addition as used in other operations, but 
here we post-normalize the result. For the  output format we use a
mantissa with the hidden bit.

\subsection{Memory interface}

The memory interface has two functions. The first one is to write the
data sent from the host, and the second one is to read the memory
during the calculation.

The data of one particle is packed into 16 32-bit words. A data packet
sent from the host consists of two control words and this 16-word
data.  The first control word contains following three fields: command
code (2 bits), chip identity (10 bits) , and chip identity mask (10
bits). The second word is the starting address in the memory for the
particle data.

The chip identity field is used to select the chip that actually
stores the particle data. With the design of GRAPE-6, all chips on one
board, or  on multiple boards connected to the same host,
receive the same data from the host. We, however, have to let
different chips calculate the forces from different particles, and
this can be achieved by specifying, in the particle data packet, the
identity of the chip that actually store the data. When a chip
receives one $j$-particle data packet, it writes the data to the
memory only if the chip identity field of the packet (masked by
the identity mask) is the same as its identity register (also masked
by the identity mask). The identity register itself must be all
different on different chips, and how we achieve this will be
discussed in the next subsection . The identity mask field is usually
all ones.

The memory interface is designed to control two SSRAM (synchronous
static random-access memory) chips with 36-bit data width. All signal
lines drive only one chip, so that we can minimize the signal
length. Using the combined data width of 72 bits, we implemented ECC
(SECDED or single error correct and double error detect) for the data
received from the memory.

The memory interface is programmable, in the sense that practically
all access latencies can be adjusted by writing to on-chip registers.
Thus, we can use almost any type of SSRAM with different access
timings.

During the calculation, both memory chips output data at every clock
cycle. The memory address counter is initialized to $8N$, where $N$ is
the number of particles, and decremented at each clock. For writing
the data, we use a slower access, where we write two SSRAM chips at
alternate clock cycles. In this way, we can reduce the switching noise
and can also relax the timing requirement for the data bus.

\subsection{I/O ports and handshake protocol}

Tables \ref{tab:jp-interface_spec} and
\ref{tab:control-pipeline-fo-external-interface_spec} show the
signal definition for input and output ports. Both ports operates on
the clock with the frequency 1/4 of that of the internal logic and memory
interface. As a result, the communication bandwidth is rather limited. 
However, the electrical design of the board is  easier with a lower
clock speed. Also, with the 32-bit data width, we can still achieve the
data transfer speed of around 100 MB/s, which is fast enough to match
with the speed of PCI bus of the host computer.

The input port is very simple, with data lines and a single write
enable line. The chip actually has two input ports, one dedicated to
the data sent to the memory (we call this the JP port), and the other
for everything else (the IP port). On the JP port, the data of one
particle consists of 18 32-bit words, and the control logic handles
this 18-word packet. The IP port is a general-purpose port. It accepts
variable-length data packet. The first word of the packet is the
starting address of the on-chip register. The second word is the
number of data words to follow, and remaining words are all data.

\begin{table}
\begin{center}
\caption{GRAPE-6 chip input port signal definition}
\begin{tabular}{lcl}
\hline
signal & width&  description \\
\hline
DATA & 36 & 32 bit data with 4 bit parity\\
WE & 1 & write enable\\
\hline
\end{tabular}
\label{tab:jp-interface_spec}
\end{center}
\end{table}

The output port is more complicated, because we need to implement flow
control. The reason we need flow control is that for some data, for
example for the neighbor list, the host must receive the data directly
from all chips. In the case of the force, all chips  output the
results synchronously and the onboard reduction network  reduces
the data on the fly. However, the neighbor list data has to be
transferred to the host without any reduction.

It is possible to read neighbor data from each chip without using
hardware flow control, by let the host computer  send the
commands  to each chip sequentially until it receives all data. In
this case, the processor chip itself does not need any flow control.
However, this procedure would be rather slow, since the host has to setup
the DMA transfer many times. Therefore, we chose to let the host to
send the command to all chips. The reduction network takes care of the
flow control. In table
\ref{tab:control-pipeline-fo-external-interface_spec}, the WD signal
is used for flow control.

When the WD signal is asserted, the chip stops sending
new data. When the chip sends a new data, it asserts both VD and
ND signals. The VD signal is asserted as long as the data is valid, but ND is asserted
only when the data is actually updated.
The STS line is a special signal which tells if the force calculation
pipeline is working or not. The ACTIVE signal is used to indicate
defective chips. The output of this pin is programmable from the host, 
and if ACTIVE is negated, the reduction network ignores the output
from the chip.

\begin{table}
\caption{FO port signal description}
\begin{center}
\begin{tabular}{lcl}
\hline
Signal & direction (I/O)&  description\\
\hline
D0-D35 & O & data (4 bits for parity)\\
VD    & O & valid data \\
ND    & O & new data \\
STS    & O & status \\
ACTIVE & O & if 0, chip is unused\\
WD    & I & wait data \\
\hline
\end{tabular}
\end{center}
\label{tab:control-pipeline-fo-external-interface_spec}
\end{table}

\section{Processor board and network hardware}
\label{sect:pbnh}
\subsection{Processor module and processor board}

\begin{figure}
\begin{center}
\leavevmode
\FigureFile(6 cm,6 cm){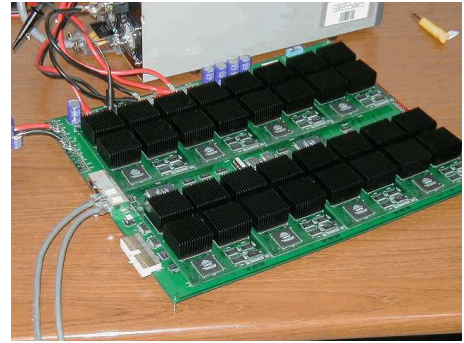}
\end{center}
\caption{The processor board.}
\label{fig:pbfig}
\end{figure}

Figures \ref{fig:processorboard} and \ref{fig:pbfig} show the
processor board. A single board houses 32 processor chips. Logically,
the design of the board is rather simple. The input data is
broadcasted to all chips, and the output data of the chips are reduced
through a reduction network.

The nodes of the reduction network are made of FPGA chips. It has two
operation modes, reduction mode and pass-through mode. In the
reduction mode, it receives the data from lower-level nodes (either
the processor chips or lower-level FPGA nodes), and performs
reduction. Since one particle data consists of force, potential, time
derivative of the force, the distance and index of the nearest
neighbor particle, and status flags, the operation of the reduction
ALU need to change according to the data type, and is controlled by a
sequencer.

In the pass-through mode, a node sends the data received from the
lower-level node without applying any operation. Since multiple
lower-level nodes might try to send the data simultaneously, every
node controls the WD signal (which is also implemented in a node FPGA
as well) so that only one chip (or node) actually sends the data at
one time. When one chip (or node) indicates the end of the data by
negating the VD signal, the node negates the WD signal for the next chip
to start receiving the data from that chip.

As can be seen in figure \ref{fig:pbfig}, one processor board is
designed to house up to eight processor ``modules''. A single module
houses 4 processor chips, 8 SSRAM chips, and an FPGA chip which
realizes the 4-input, 1-output reduction tree.  We made this division
between the board and module, to make the manufacturing easier. With
this separation, all BGA chips (with large number of pins) are mounted
on small-size module boards. Thus, the rate of the soldering error
should be lower, compared to the case where we mount them on large
boards. In addition, if there is an error, only a module with 4 chips
would be defective. Of course, having to connect the board and module
through a connector increases the probability of the failure, but we
expected that the failure rate of the connector is significantly lower
than the failure rate of the soldering (which turned out to be the
case)

The tree nodes are implemented using Altera ACEX series chips. In the
lowest level (processor module level), we used EP1K50A chips in
484-connect BGA packages. This chip implements a four-input node.
Higher levels are implemented on EP1K30A chips in 208-pin QFP
packages. This chip implements a 2-input node. These nodes are on the
processor board.

The processor board is an 8-layer standard PCB. The processor module
board is an 11-layer board with inner via holes. The FPGA and processor 
chips are mounted on the top side, and SSRAM chips on the bottom
side. By this layout, we can minimize the wire length between SSRAM
chips and processor chips, and still achieve rather high packaging
density. We used 4Mbit SSRAM chips. Two SSRAM chips connected to a
processor chip can store up to 16,384 particles. One board can
store up to 524,288 particles.

The SSRAM chips we chose requires 2.5V power supply for I/O and 3.3V
for core. Both the processor board and module board have separate
power planes for both 2.5 and 3.3V power supply.

Though the chip has separate ports for $j$-particles and other data,
for the board we decided to use a common data line, to simplify the
design and reduce the manufacturing cost.

Currently, the core of the processor chip operates on a 90 MHz clock,
and the I/O part on a 22.5 MHz clock. the reduction network and other
logics of the control board operate also on the 22.5MHz clock.

For board-board connection, we used a semi-serial LVDS signal. We used
4-wire (3 for signals and 1 for transmission clock) chipset, which
performs 7:1 parallel-serial conversion. Since our basic transfer unit
is a 32-bit word, we used two cycles of this chipset to transmit one
data. Thus, the chipset operates on a 45 MHz clock, and the signal
lines operate at the data rate of 315 MHz. For the conversion between
the 22.5 MHz data rate of the board logic and the 45 MHz data rate of
the LVDS chipset, we used additional FPGA chip.

With this LVDS chipset, the receiver chipset itself is driven by the
clock signal which comes with the data.  In order to allow the two boards
connected to a link to operate on independent clocks, we added FIFO
chips after the data rate is reduced to 22.5 MHz. 

The physical form factor of the card is that of an 8U Eurocard (with the
length of 400mm). For the backplane connection, we used connectors
designed for Compact PCI cards. The power supply is also from backplane bus,
through special power connectors.

It is possible to connect a single processor board directly to the
host through the host interface card, without using the network
card. For this purpose, the processor board also has the connector for 
the twisted-pair cable for the LVDS interface. These connectors are
standard RJ-45 modular jacks  widely used for 10/100/1000BT
Ethernet connection. Standard category 5 (or enhanced 5)
cables can be used for connection.

For LVDS interface chips, we used SN75LVD85 and  SN75LVD86A chips from
Texas Instruments.

\subsection{Host interface card}

\begin{figure}
\begin{center}
\leavevmode
\FigureFile(\columnwidth,6 cm){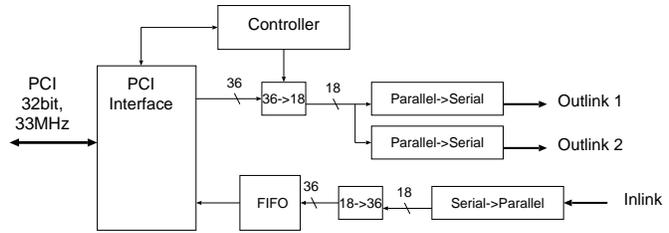}
\end{center}
\caption{The host interface board.}
\label{fig:hostinterface}
\end{figure}

Figure \ref{fig:hostinterface} shows the block diagram of the host
interface card. It is a standard (32-bit, 33 MHz) PCI card. To
transfer the data from host to GRAPE-6, the host setup the data to be
transferred in its memory and let the PCI interface chip on the
interface card  perform DMA transfer. The data received by this DMA
transfer is sent directly through the output link. In the design of
the host interface card we implemented two output ports so that they can
separately supply data to the JP and IP ports.  As stated earlier, we
decided to use only one port for the processor board. Therefore, the
second output port of the interface card is not used.

The input port is more complicated, with an FIFO memory to store the
received data. This FIFO memory is necessary, since we cannot
guarantee the response time of the host operating system to the DMA
request from the interface card. We need to have the memory large
enough to avoid any possible overflow.

For the PCI interface, we used the 9080 chip from PLX technology.

\subsection{Network board}

\begin{figure}
\begin{center}
\leavevmode
\FigureFile(6 cm,6 cm){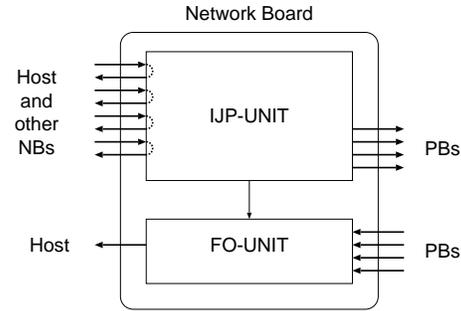}
\end{center}
\caption{The network board.}
\label{fig:cboverall}
\end{figure}

Figure \ref{fig:cboverall} shows the block diagram of the network
board. It has two basic functions. One is to broadcast (or multicast)
the data received from the host (or possibly higher-level network
boards) to the processor boards (or lower-level network boards). This
part is shown as IJP-UNIT. The other is the reduction network for the
calculated result, shown as FO-UNIT. The reduction network is exactly
the same as that on the processor board, except for the fact
that the interface to the module board is replaced by the interface
for the processor board (with LVDS link chipset).

The IJP-UNIT has 4 input ports. One of them is a special port designed
to connect to the host. Other three ports are designed to accept data
from other network boards (see figure \ref{fig:g6cluster}). Each of
the four input ports has a ``copy'' output, shown in the left-hand
side of the unit,  so that we can cascade multiple network boards.

\begin{figure}
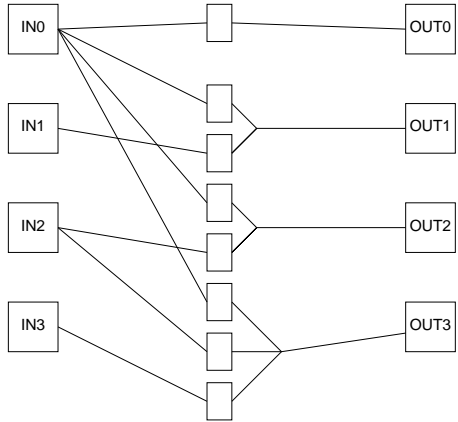

\begin{center}
\leavevmode
\FigureFile(6 cm,6 cm){cboard_jp_logic2.eps}
\end{center}
\caption{Physical implementation of multicast network.}
\label{fig:cbjp}
\end{figure}

Figure \ref{fig:cbjp} shows the block diagram of the multicast
network. The boxes in the center of the figure are all data buffers
with output enable control input, which realize the multicast
network. Note that this structure implements the network 
logically equivalent to what is shown in figure \ref{fig:scalablenet}.

The control input for these buffers are supplied from the control
logic implemented on the FPGA for 45MHz-22.5 MHz data rate change.
This FPGA integrates a sequencer to decode IP/JP port data packets,
which reacts to the address space assigned to the network board.

%
% 20030328
%
\subsection{Packaging and Power distribution}

In the standard configuration, eight processor boards and two network
boards are installed in a card rack with a special backplane for the
LVDS link.  The network board is a single-height unit, but the processor
board occupies two-unit height, to allow sufficient airflow.

To both of network boards and processor boards, the electrical power
is supplied through backplane connectors. However, in our present
packaging, each processor board has its own power supply unit. A power
supply unit accepts DC 330V input, and supplies DC 2.5V and 3.3V. The
DC 330V power is generated by another power unit from three-phase AC
power line. For all these power units, we used products from Vicor.

We chose Vicor product primarily to reduce the response time of the
power supply to the change in power consumption by the boards. One
advantage of the CMOS logic is that it consumes power only when the
logic state changes. This means that even though we had paid
absolutely no effort to reduce the power consumption of the chip, its
power consumption almost halves when the pipeline are not active.

This ``feature'' of the chip is rather good from the point of view of
the running cost of the machine, but pauses a rather serious problem to
the power supply. The typical response timescale of a switching power
supply units is of the order of one millisecond. On the other hand,
GRAPE-6 switches between calculation and idle (or communication) states
also in about one millisecond. This means that the response time of the
power supply is too long to compensate for the change in the load
between the calculation state and the idle state, and the supply
voltage becomes rather unstable. Thus, we had to look for power
supplies with a relatively short response time. For switching power
supplies, a short response time means high operating frequency, and
Vicor products had the highest frequency among commercially available
power units.

Even with high-frequency power supplies, the response time was still
the order of 100 microseconds, and the only way to stabilize the power
supply is to add large bypass capacitors. We attached capacitors with
total capacitance of about 0.1F to the 2.5V power line of each
processor board. We could not use usual alminium electrolytic
capacitors because their internal resistance (equivalent series
resistance, ESR) is too large. We used low-ESR electrolytic capacitors
from Sanyo to meet our need.

In hindsights, it would be probably better to design a small switching
power supply unit integrated into the processor module, since such a
power supply unit, which is used on every motherboard for PCs, is
inexpensive and highly reliable.

\begin{figure}
\begin{center}
\leavevmode
\FigureFile(6 cm,6 cm){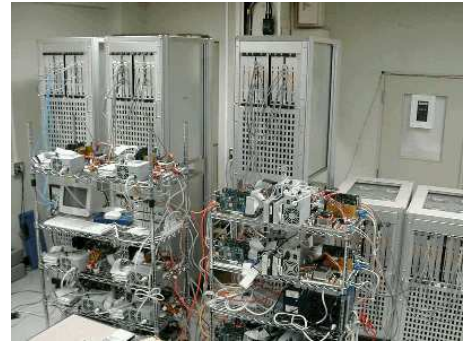}
\end{center}
\caption{The 64-board, 4-cluster GRAPE-6 with the racks for host
computers in front.}
\label{fig:g6photo}
\end{figure}

Figure \ref{fig:g6photo} shows the complete GRAPE-6 system consisting
of five racks (three with two subracks and two with one subracks),
with 16 host computers in front of them. Host computers are
Linux-running PCs, with AMD Athlon XP 1800+ processors and ECS K7S6A
motherboards. They are connected with Gigabit Ethernets.  The total
power consumption of the system is around 40 KW, when in full
operation. 

\section{Differences between GRAPE-4 and GRAPE-6}
\label{sect:g4g6}

As described in the previous sections, the architecture of GRAPE-6 is
quite different from that of GRAPE-4, even though it is the
direct successor of GRAPE-4 for essentially the same goal. In this
section, we describe what design changes are made why.

\subsection{Differences in the semiconductor technology}

The primary difference is that for GRAPE-6 processor chip we used
$0.25 {\rm \mu m}$ design rule, while with GRAPE-4 we used 
$1 {\rm \mu m}$ design rule. This difference with additional advance
in wiring enables us to integrate roughly 20 times larger number of
transistors, with 3-4 times faster clock speed. Thus,  roughly
speaking, a single GRAPE-6 chip offers the speed two orders of
magnitude higher than that of GRAPE-4.

This large advance, however, implies almost every design decision had
to be changed. In the following, we summarize the changes made.

\subsection{The host computer and overall architecture}

In  GRAPE-4, 4 clusters are connected to a single
host, sharing one I/O bus. For the peak speed of 1 Tflops, the
single host was still okay for simulations with large number of
particles ($10^5$ and larger), and communication through a single I/O
bus was also okay.

With GRAPE-6, however, the peak speed is increased  by a factor
of 60. On the other hand, the speed of a single host would be
improved only by a factor of 10 or so, if we assume the standard
Moore's law (performance doubling time of 18 months). Thus, if we want
to achieve a reasonable speed for similar number of particles as that
for GRAPE-4, we need to use around 10 host computers and the
communication channel must be 10-20 times faster than that  used
for GRAPE-4.

Around the time of the design, it was clear that a shared-memory
multiprocessor system with 8-16 processors and sufficient I/O
bandwidth would be prohibitingly expensive, with the price tag of the
order of 1 M USD. On the other hand, a cluster of 8-16
single-processor workstations or PCs would be much less expensive. As
far as the cost is concerned, clearly a cluster of single-processor
machines was better than a shared-memory multiprocessor system.

\begin{figure}
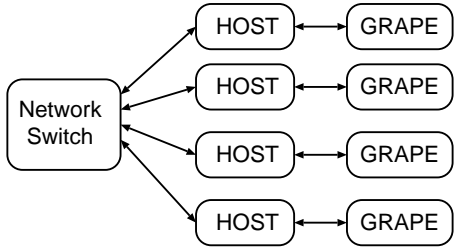

\begin{center}
\leavevmode
\FigureFile(6 cm,6 cm){simpleparallelhost.eps}
\end{center}
\caption{A simple parallel-host, parallel-GRAPE system.}
\label{fig:simpleparallel}
\end{figure}

One problem with the cluster is that the simplest configuration (see
figure \ref{fig:simpleparallel}) does not work. The reason is the
following.

With this configuration, there are two different ways to distribute
particle data over processors\citep{Makino2002}. One is that each
processor has the complete copy of the system (the ``copy''
algorithm). In this case, parallelization is performed as follows. At
each blockstep, each processor determines which particles it
updates. After all processors update their share of particles, they
exchange the updated particles so that all processors have the updated
copy of the system. This algorithm has been used to implement the
individual timestep algorithm on distributed-memory parallel computers
\citep{SpurzemBaumgardt1999}

In this algorithm, at the end of the block timestep each processor
receives the particles updated on all other processors. This means the
amount of communication is independent of (or, strictly speaking, is a
slowly increasing function of) the number of processors, and the
overall performance of the system is limited by the speed of
communication.

The other possibility is to let each processor to have a
non-overlapping subset of the system, so that one particle resides
only in one processor. In this case, with the blockstep algorithm we
need to pass around the particles in the current blockstep, so that
each processor can calculate the forces from its own particles to
particles on other processors(the ``ring'' algorithm). The amount of
communication (host-host and host-GRAPE) per blockstep is again
independent of the number of processors. This algorithm is also
implemented on distributed-memory parallel computers with direct
summation \citep{Dorbandetal2003} and even with the tree algorithm
\citep{Springeletal2000}.

For general-purpose parallel computers, this simple algorithm actually
works rather well, simply because the calculation speed of single node
is so slow. Even a cluster with several hundred nodes is still slower
than a single GRAPE-4. So the communication speed of 10-100 MB/s is
sufficient. However, with GRAPE-6 we do need a faster speed.

Now we understand that it is possible to use a hybrid of the above two
algorithm to solve the bottleneck \citep{Makino2002}. In this hybrid
algorithm, we organize processors into two-dimensional grid, and
distribute the particles so that each row (and each column) has the
complete copy of the system.

In the standard realization, this algorithm requires that total number
of processors is $r^2$, where $r$ is a positive integer number. We
divide $N$ particles into $r$ subsets, each with $N/r$ particles.  If
we number processors from $p_{11}$ to $p_{rr}$, processor $p_{ij}$ has
the copy of both $i$-th and $j$-th subsets.

At the beginning of the each blockstep, each processor selects the
particles to be updated from subset $i$. Then all of them calculate
the force on them from subset $j$. After that, the total forces can be
calculated by taking summation over columns. Here, we assume the
summed results are obtained on diagonal processors $p_{ii}$. 

After particles in the
current block are updated on $p_{ii}$, they are broadcasted
to all other processors in the same row ($p_{xi}$) and also in the
same column ($p_{ix}$)
so that both subsets $i$ and $j$ are updated on each processor.

In this algorithm, the amount of communication for one node is
$O(N/r)$. In other words, the effective communication bandwidth (both
host-host and host-GRAPE) is increased by a factor $r$. Thus, the
communication speed is improved by a factor proportional to the square
root of the number of processors.

At present, this solution looks fine, since the price of the fastest
single-processor frontend is now rather cheap. The cost of the
communication is also rather cheap, with Gigabit Ethernet adapters
available for less than 100 USD per unit.

When we started the design of GRAPE-6 in 1996, we did not expected
such a drastic change in the price of fast frontend processors. At
that time, RISC microprocessors were still several times faster than
PCs with IA-32 architecture, and 100Mbit Ethernet adapters were still
expensive. Thus, we had to come up with a design that did not need
$r^2$ processors or fast host-host communication.

\begin{figure}
\begin{center}
\leavevmode
\FigureFile(6 cm,6 cm){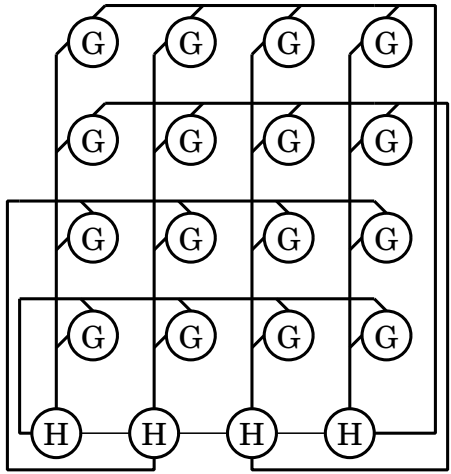}
\end{center}
\caption{Two-dimensional network of GRAPE hardwares connected to
one-dimensional host network.}
\label{fig:g6net}
\end{figure}

It was not really difficult to come up with such a design, since the
only thing non-diagonal processors does is the force
calculation. Instead of two-dimensional grid of host processors, we
can construct a two-dimensional grid of GRAPE hardwares with
orthogonal broadcast networks (figure \ref{fig:g6net}). The GRAPE
hardwares in the same row store the same data to their particle
memories. When they calculate the forces, GRAPEs in the same column
receive the same particles and calculate forces on them from particles
in the memory. The calculated results on boards in the same column are
then summed and returned to the host.

One practical problem with this network architecture is that we cannot
divide the system to smaller configurations so that we can run
multiple programs. In the case of $r^2$ hosts, we can divide the
system to any sub-squares, down to $r^2$ single host-GRAPE pairs. In
the case of 2D hardware network, we do not have any such
division. This problem can be partly circumvented by attaching a
simple switching network before memory interface, so that they can
select input. So we adopted the network structure shown in figure
\ref{fig:scalablenet}.

In the final design of GRAPE-6, we actually adopted a hybrid of
host-grid approach and GRAPE-network approach, to make a reasonable
compromise between the flexibility and absolute performance. Of
course, this shift from the pure hardware network to hybrid one is
made partly because we took into account the evolution of the host
computers during the development period of GRAPE-6. It has become more
cost effective to use large number of inexpensive (yet fast)
computers as host than to have an elaborate hardware network to
connect GRAPEs to small number of hosts.

\subsection{board-board connection}

GRAPE-4 consisted of 36 processor boards, organized in a two-stage
simple tree network. Nine boards are housed in one rack, with one
backplane bus. These boards are all connected to a control board,
which broadcasts the data from the host to all processor boards and
take the summation of the calculated data on nine processor
boards. Since all boards are connected through a shared backplane bus,
the control board has to access processor boards sequentially. In
order to improve the data transfer rate, we used a wide data bus with
the width of 96 bits.

The connection between the control board and the host was a 32-bit
parallel connection through a coaxial flat cable. This connection is
robust and reliable, but had three drawbacks: it was physically large,
it was difficult to use long wires, and it was pretty
expensive. Because a common clock signal is used on the both side of
the connection, the wire length is limited by the allowable signal
skew, which means it is difficult to use fast clock (GRAPE-4 used 16
MHz clock).

A more practical problem is that board-board wiring would become too
bulky and cumbersome, with hundreds of flat cables and nearly 10,000
contact points, if we use the same connection for GRAPE-6.  In
particular, it would be  difficult to design the network board, since it
needs to have more than 10 connectors. Also, it would be impractical
to use backplane to connect the network board and processor boards,
since the number of pins on the network board would be too large.

An obvious solution for this problem is to use a fast serial signal,
such as the physical layer of the Gigabit Ethernet. At the time of our
design decision, however, Gigabit Ethernet was unpractical because
copper wire connection was not available in 1998. Optical
connection would be too expensive and would dissipate too much heat.

We adopted what is called ``LVDS Link'' or ``Flat Panel Display (FPD)
link'', which uses four twisted-pair differential signal lines (three
for signals and one for clock). The reason we chose this interface was
that inexpensive serializer/deserializer chips were commercially
available and that we could use standard category 5 shielded 4-pair
cables for 100Mbit Ethernet cable and its connectors for reliable data
transmission, for the cable length of up to about 5 meters.

Additional advantage of this choice is that we can use backplane
connection (with custom-designed signal pattern) for connection
between the network board and processor boards. Because the number of
signals is small (8 for one port), we can pack many ports into a
standard backplane connector (we adopted Compact PCI connector).

\subsection{Pipeline chip and memory interface}

The processor chip for GRAPE-4 had a single pipeline, which calculates 
the force on two particles in every six clock cycles (2-way
VMP). During force calculation, the chip receives the data of one
particle (position, velocity and mass) in every three external clock
cycles, and the width of the input data bus was 107 bits.

One GRAPE-4 board housed 48 pipeline chips, all of which receive the
same particle data from the memory and calculate the force on two
particles. This means that a single board calculates forces on 96
particles in parallel.

This shared-memory architecture is simple to implement. However, we
could not use this architecture for GRAPE-6, since the hardware
parallelism would become excessively large. The pipeline chip for
GRAPE-6 would be roughly 50 times faster than that for GRAPE-4. Thus,
even if we somehow increase the data transfer rate by a factor of 5,
the number of particles on which the forces are calculated in parallel 
would increase by a factor of 10, from 100 to 1,000. This number is
too large, if we want to obtain a reasonable performance for
simulations of star clusters with small, high-density cores. Note that
with multiple-board configurations, this number would become even
larger. On an $r\times r$ two-dimensional system, the degree of
parallelism becomes larger by a factor of $r$. 

The data transfer rate of GRAPE-4 chip was about 200 MB/s. To keep the
degree of parallelism to be around 100 or less, the GRAPE-6 chip would
have to have the data transfer rate of 5 GB/s, which was well beyond
our capability of designing and manufacturing. At 100 MHz clock, the
speed of 5 GB/s requires 400 input pins. It is quite difficult to have
400 signal lines, all with 100 MHz data rate, to connect more than a
few chips.

Clearly, a different design was necessary. Too large degrees of
parallelism arose from our decision to let a large number of chips to
share one memory unit. If we reduce the number of chips to share the
memory, thus, we can solve the problem. The extreme solution is to
attach one memory unit to each pipeline chip, and let multiple
pipelines to calculate the force on the same set of chips, but from
different set of particles.

This extreme solution has one important practical advantage. 
The connection between the processor chip and its memory is
point-to-point, and physically short (since we can put a processor chip 
and its memory next to each other). This means a high clock
frequency, such as 100 MHz, is relatively easy to achieve.

To attach memory chips directly to the processor chips, we need to
integrate the predictor pipeline and the memory controller unit
(generation of address and other control signals) to the processor
chip. These do not consume much transistors. Therefore it does not
have any effect to  the performance of the chip. 

With GRAPE-6, we adopted a 72-bit (with ECC) data width for transfer
between memory and the processor chip. A GRAPE-6 chip integrates six
8-way VMP pipelines. Therefore it calculates the forces on 48
particles in parallel. All pipelines on board calculate the
forces on the same set of particles. Thus, even with the largest
configuration we considered (an $8 \times 8$ system), the degree of
the parallelism is still less than 400, not much different from that
of full-size GRAPE-4 (which was also 400).

This change from shared memory design to local memory design implied
we had to take summation of large number of partial forces obtained
on chips on one board. With GRAPE-4, we also had to take summation of
forces obtained on different boards, and we used commercially available
single-chip floating-point arithmetic units for this summation. With
GRAPE-6, we could not apply this solution simply because such chips no 
longer existed. Thus, we have to either integrate this summation
function into the processor chip, or develop another chip to take
summation.

We adopted the latter approach, but used FPGA (Field-programmable gate
array) chips to implement adders. It was not impossible to integrate
floating-point adders into FPGAs, but such a design would require
rather large, expensive FPGA chips and a complex design. In order to
simplify the design, we chose to use a block floating point format for
the force and other calculated result. In this format, we specify the
exponent of the result before we start calculation. The actual value
of exponent can be different for forces on different particles, so
that we can calculate the forces with wildly different magnitudes in
parallel.

With this block floating point method, we can greatly simplify the
design of the hardware to take the summation. Of course, we have to
supply the value of exponent, but the value of the exponent at the
previous timestep is almost always okay. For the initial calculation,
we sometimes need to repeat the force calculation a few times until
we have a good guess for the exponent.

A rather important advantage of using the block floating point format
is that the calculated result is independent of the number of
processor chips used to calculate one force. Since the actual
summations, both within the chip and outside the chip, are done in
fixed-point format, no round-off error is generated during
summation. Of course, round-off error is generated when we shift the
calculated force to meet the block floating point format, but this
error is independent of the order in which the summation is
performed. In the case of the usual floating-point format used in
GRAPE-4, the round-off error generated in the summation depends on the
order in which the forces from different particles are accumulated,
and therefore the calculated force is not exactly the same, if the
number of boards in the system is different.

Of course, this difference does not have any effect on the accuracy of 
the simulation itself, since the word length itself is chosen as
such. However, it is quite useful to be able to obtain exactly the same 
results on machines with different sizes, since it makes the
validation of the result much simpler.

\section{Performance}
\label{sect:performance}

In this section, we discuss the performance of GRAPE-6 system, both
for the direct summation algorithm with individual timestep and the
tree algorithm.  For both algorithms, we discuss the performance of
single-host system and multi-host system.

\subsection{Direct summation with individual timestep}

Here we discuss the performance of GRAPE-6 for the individual timestep
algorithm. As the benchmark run, we integrate the Plummer model with
equal-mass particles for 1 time unit (we use the ``Heggie'' unit, \cite{HeggieMathieu1986},
where the gravitational constant $G$ and total mass of the system $M$
are both unity and the total energy of the system $E$ is $-1/4$). We
used standard Hermite integrator \citep{MakinoAarseth1992} with the
third-order predictor. Timestep criterion is that of Aarseth
\citep{Aarseth1999b} with $\eta = 0.01$. For softening parameter, we
tried three different choices. The first one is a constant softening,
$\epsilon=1/64$. We also tried $\epsilon = 1/[8(2N)^{1/3}]$ and
$\epsilon = 4/{N}$, to investigate the effect of the softening
size. Note that for $N=256$, all three choices of the softening give
the same value. In the following, we first describe the performance of
a single-host system (with 4 processor boards). Then we discuss the
performance of a single cluster with 2 or 4 hosts, and finally we
discuss the performance of multiple-cluster configurations.

\subsubsection{single-host performance}

\begin{figure}
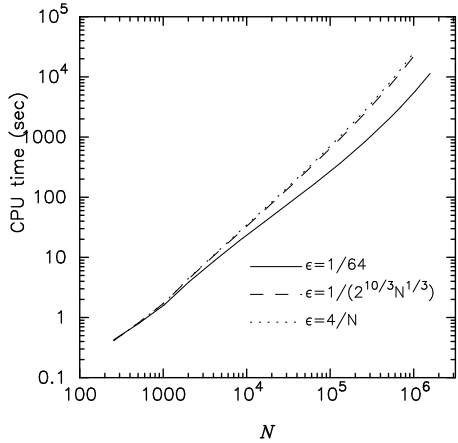

\begin{center}
\leavevmode
\FigureFile(6 cm,6 cm){singletime.ps}
\end{center}
\caption{CPU time in second to integrate a Plummer model for 1 time
unit plotted versus the number of particles $N$. Solid, dashed and
dotted curves indicate the result with constant,
$1/{N}^{1/3}$ and $1/N$ softenings, respectively.}
\label{fig:singletime}
\end{figure}

Figure \ref{fig:singletime} shows the CPU time to integrate the system
for one time units. We actually measured the CPU time for integration
from time 0.25 to 1.0 and multiplied the result by 4/3, since the step
size after the start of the integration is too small because of the
initialization procedure. From figure \ref{fig:singletime} we can see
that the CPU time is almost proportional to $N$ for $N<10^5$, but for
$N$-dependent softenings the dependence is slightly higher. For
$N>10^5$, the slope approaches to 2.

\begin{figure}
\begin{center}
\leavevmode
\FigureFile(6 cm,6 cm){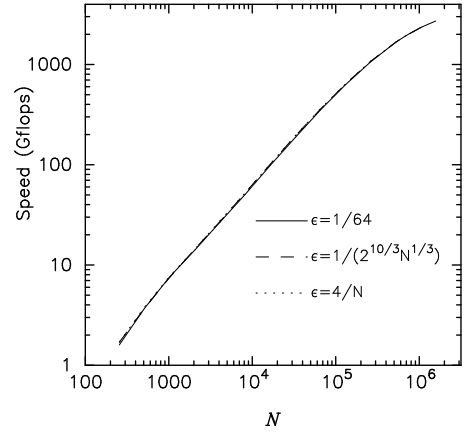}
\end{center}
\caption{Same as figure \protect \ref{fig:singletime}, but calculation 
speed in Gflops is plotted.}
\label{fig:singlespeed}
\end{figure}

Figure \ref{fig:singlespeed} shows the actual calculation speed
achieved. The theoretical peak speed of the single-host, 4-PB system
is 3.94 Tflops.  Here, we define the calculation speed as
\begin{equation}
S = 57N n_{steps},
\end{equation}
where $n_{steps}$ is the average number of individual steps performed
per second. The factor 57 means we count one pairwise force
calculation as 57 floating-point operations. We took this number  from
recent literatures. From this figure, we can 
see that the achieved speed is practically independent of the choice
of the softening. The reason why calculations with smaller softening
takes more CPU time is that the number of timesteps is larger, as
shown in figure \ref{fig:singlesteps}. For calculations with
$N$-dependent softenings, the number of block steps increases
significantly as we increases $N$. This means that the average number of
particles in one block grows rather slowly. However, as we can see
from figure \ref{fig:singlespeed} this does not affect the achieved
performance.

\begin{figure}
\begin{center}
\leavevmode
\FigureFile(6 cm,6 cm){singlesteps.ps}
\end{center}
\caption{Same as figure \protect \ref{fig:singletime}, but the number
of total
individual steps (upper) and block steps (lower) are plotted. }
\label{fig:singlesteps}
\end{figure}

Roughly speaking, we can model the calculation time per one particle
step as follows:
\begin{equation}
T_{single} = (1-f)T_{host}+T_{comm}+\max(T_{GRAPE},fT_{host}),
\end{equation}
where $T_{host}$ is the time for the host computer to perform
computations  to integrate one particle, $T_{comm}$ is the time
needed for the communication, and $T_{GRAPE}$ is the time to calculate 
the force on GRAPE. The factor $f$ is the fraction of the operations that
the host computer can perform while GRAPE is calculating the force. The
program we used tries to perform the time integration on host and the force
calculation on GRAPE with as much concurrency as possible.

%This is clearly oversimplified, since $T_{host}$ weakly depends on
%$N$. The reason is that for small $N$ the cache-hit rate is higher and 
%thus $T_{host}$ is somewhat smaller.

We can estimate $T_{comm}$ as follows. The total amount of data
transferred for one particle step is currently 200 bytes. With the present
host, the effective data transfer rate for DMA transfer is 80
MB/s. Therefore
\begin{equation}
T_{comm}= 200/(8\times 10^7) = 2.5\times 10^{-6} {\rm sec}.
\end{equation}
The calculation time on GRAPE is expressed as
\begin{equation}
T_{GRAPE}= N/(9\times 10^7 n_{pipes} ) = 1.447\times 10^{-11}N {\rm sec},
\end{equation}
where $n_{pipes}$ is the total number of pipelines. With our current
system $n_{pipes}=768$.

\begin{figure}
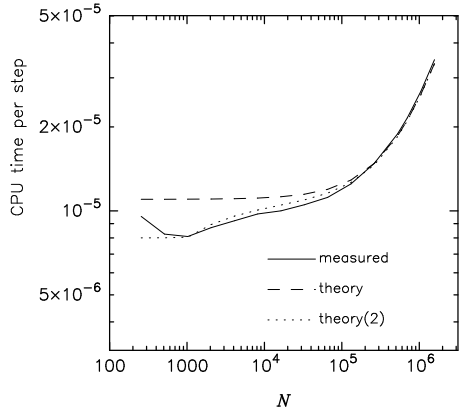

\begin{center}
\leavevmode
\FigureFile(6 cm,6 cm){singletheory.ps}
\end{center}
\caption{CPU time per one particle step plotted as a function of the
number of particles $N$. Solid curve is the measured result. Dashed
and dotted curves denote two different theoretical estimates. }
\label{fig:singletheory}
\end{figure}

In figure \ref{fig:singletheory}, the solid curve shows the measured
CPU time per step. The dashed curve is a fit, with
$T_{host}=8.5\times 10^{-6}{\rm sec}$ and $f=0$. We can see that
agreement between the theory and experimental result is good for large 
$N$, but is rather poor for small $N$. This is
because we ignored the effect of the cache memory on $T_{host}$. The
dotted curve is the theoretical estimate with a heuristic model for
the cache effect. For this curve, we used
\begin{equation}
T_{host } = 5.5\times 10^{-6} c +  8.5\times 10^{-6} (1-c) (\rm{sec}),
\end{equation}
where $c$ is expressed as
\begin{equation}
c = \cases{
         1, &$(N\le 1000)$,\cr
         \sqrt{N/1000},&$(N>1000)$.
}
\label{eq:cache}
\end{equation}
This model is purely empirical, but apparently gives a reasonable
description for the performance. Since this effect of the cache is
rather large, it turned out to be difficult to determine the value of
$f$ empirically. We assumed $f=0$.

For $N<1000$, the experimental value is larger than the prediction of
the refined theory. This is because the number of particles in one
block is too small. The overhead to invoke DMA operations becomes
visible.

\begin{figure}
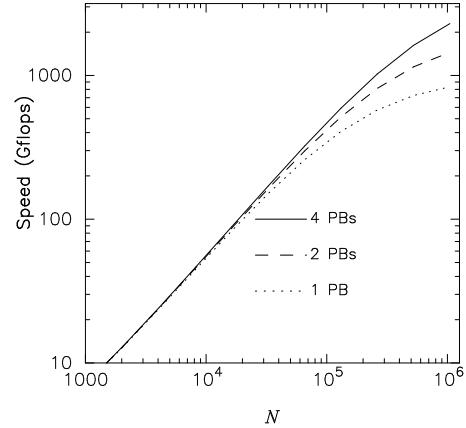

\begin{center}
\leavevmode
\FigureFile(6 cm,6 cm){singlespeedtheory.ps}
\end{center}
\caption{The estimated performance of 4-, 2- and 1-PB systems
as the function of the number of particles $N$. Solid, dashed and
dotted curves denote the speed of 4-, 2- and 1-PB systems,
respectively.  }
\label{fig:singlespeedtheory}
\end{figure}

Up to here, we discussed only the speed of a 4-PB system. Since there
are many installation of GRAPE-6 outside Tokyo university with one
PB connected to a host, it would be useful to give the performance of
smaller configurations. 
Figure \ref{fig:singlespeedtheory} gives the estimated performance of
4-, 2- and 1-PB system. One can see that performance difference is
rather small for $N<3\times 10^4$. For $N>10^5$, performance
difference becomes significant. 

\subsubsection{multi-host performance}

\begin{figure}
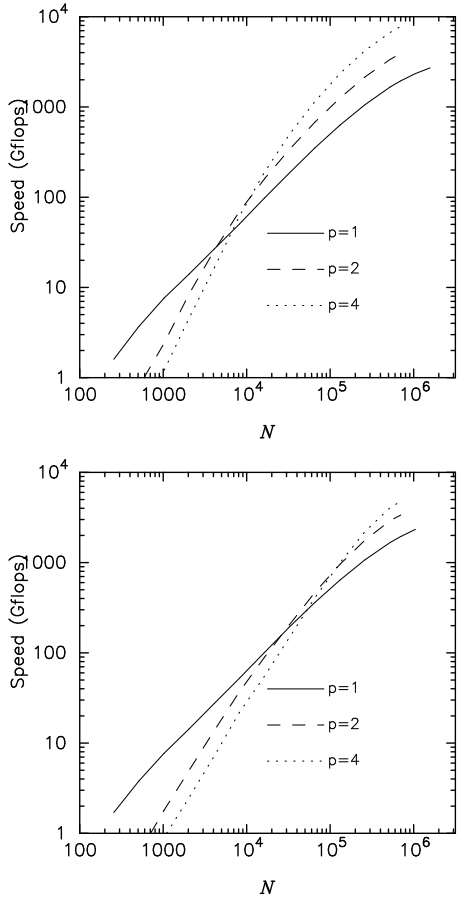

\begin{center}
\leavevmode
\FigureFile(6 cm,6 cm){np4speed.ps}
\FigureFile(6 cm,6 cm){np4speed3.ps}
\end{center}
\caption{The  calculation speed in Gflops plotted as a function of
$N$. Solid, dashed and dotted curves show the results for 1, 2 and
4-node systems, respectively. The left panel shows the result for
constant softening, and the right panel $\epsilon=4/N$.}
\label{fig:np4speed}
\end{figure}

Figure \ref{fig:np4speed} shows the calculation speed for multi-host
systems with up to 4 hosts. The peak speed of 2- and 4-hosts systems
are 7.88 Tflops and 15.76 Tflops, respectively. For up to 4 hosts, the
network boards are used to distribute the data, and the communication
network between the host computers are used primarily for
synchronization. The parallel program itself is written using MPI, and
we used MPICH/p4 over TCP/IP as the MPI library. The network interface
is Planex GN-1000TC Gigabit NIC, which uses NS 83820 chip. We found
the performance of MPICH/p4 on this network interface to be quite
unsatisfactory, and used UNIX TCP/IP socket system calls for actual
communication.

We can see that multi-host codes require rather large number of
particles to achieve the speed faster than that of the single-host
code. Even with the constant softening, the two-host code becomes
faster than the single-host code only at $N\sim 3000$, and for
$\epsilon = 4/N$, this crossover point moves to around $N\sim 
10^4$.

\begin{figure}
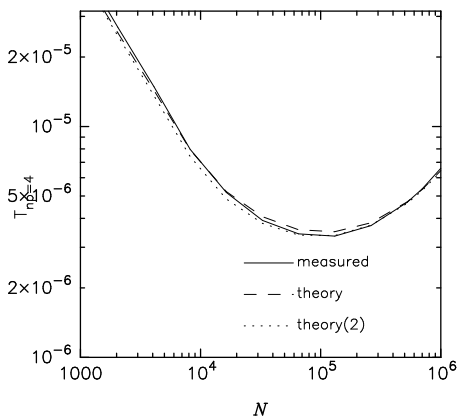

\begin{center}
\leavevmode
\FigureFile(6 cm,6 cm){np4theory.ps}
\end{center}
\caption{Same as figure \protect \ref{fig:singletheory} but for the
case of 4-node parallel calculation.}
\label{fig:np4theory}
\end{figure}

Figure \ref{fig:np4theory} shows the calculation time per one particle 
step for 4-node parallel calculation. The measured value is obtained
by dividing the total number of particle steps by the wallclock
time. This figure clearly shows why the value of $N$ for the crossover 
is rather large. For ``small'' $N$ ($N<10^4$), the calculation time is 
inversely proportional to the number of particles $N$. This is because 
the communication between hosts, which takes constant time per
one blockstep, dominates the total cost in this regime. To be
quantitative, the calculation time per one particle step is expressed
as
\begin{equation}
T_{mn,p} = T_{single}/p + T_{comm,hosts},
\label{eq:Tmnp}
\end{equation}
where $p$ is the number of nodes in a cluster and $T_{comm,hosts}$ is
the communication time expressed as
\begin{equation}
T_{comm,hosts} = 6(\log_2p+1 )t_{sync}/n_b,
\end{equation}
where $(\log_2p +1)t_{sync}$ is the time to complete a barrier
synchronization for parallel code running on $p$ nodes. The
logarithmic factor comes from the fact that synchronization requires
$\log_2p+1 $ stages. The divisor, $n_b$, is the average number of
particles integrated in one blockstep. For our current implementation
of the synchronization, we found $t_{sync} = 250{\rm \mu s}$. The
factor 6 is the number of synchronization operations necessary
in one blockstep.

Theoretical estimates shown in figure \ref{fig:np4theory} are
calculated using equation (\ref{eq:Tmnp}). Here again, the agreement
between the measured result and the theory with the effect of the
cache memory of the host is very good. To evaluate
$T_{host}$, we used $N/p$ instead of $N$ in equation (\ref{eq:cache}), 
since one node handles $N/p$ particles.

\subsubsection{multi-cluster performance}

\begin{figure}
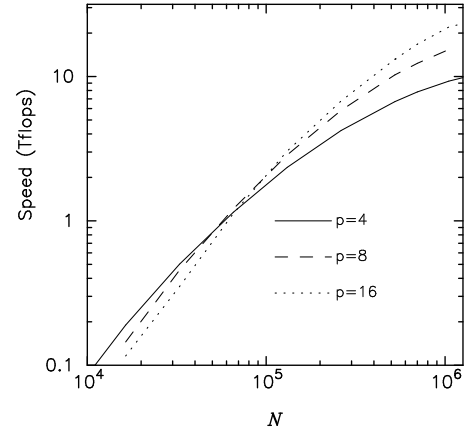

\begin{center}
\leavevmode
\FigureFile(6 cm,6 cm){np16speed.ps}
\end{center}
\caption{The  calculation speed in Tflops plotted as a function of
$N$. Solid, dashed and dotted curves show the results for 4, 8 and
16-host (1, 2, and 4-cluster )systems, respectively. Constant
softening is used for all runs.}
\label{fig:np16speed}
\end{figure}

Figure \ref{fig:np16speed} shows the calculation speed for
multiple-cluster systems, as a function of the number of particles in
the system $N$. The crossover point at which multi-cluster systems
becomes faster than single-cluster system is rather high ($N \sim
10^5$), and even for $N=10^6$, the speedup factors achieved by
multi-cluster systems are significantly smaller than the ideal
speedup.

For multi-cluster system, the calculation time per one particle step
can be estimated as follows. In our current implementation of the
multi-cluster calculation code, one host of a $p$-hosts, $q$-cluster
system (therefore $p/q$ hosts in a cluster) handles $N/p$
particles. The forces from the particles in the hosts in the same
cluster can be calculated using the hardware network on the side of
the GRAPE-6. However, one cluster need to gather the information of
particles on different clusters.  By letting each
of $p/q$ hosts in one cluster  receive data from other $q-1$ hosts,
we can let one cluster  maintain complete date of  all $N$
particles. This is just one of many possible implementations. For
small value of $q$, theoretically,  this is close to the best possible
implementation. 

\begin{figure}
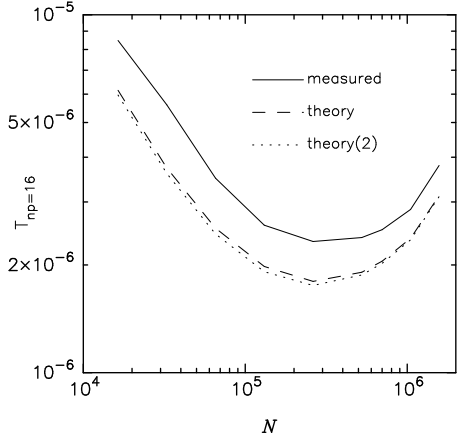

\begin{center}
\leavevmode
\FigureFile(6 cm,6 cm){np16theory.ps}
\end{center}
\caption{Same as figure \protect \ref{fig:singletheory} but for the
case of 16-host parallel calculation.}
\label{fig:np16theory}
\end{figure}

With this implementation, the calculation time per one particle step
is expressed as
\begin{equation}
T_{mc,p} = T_{single}/p + T_{comm,hosts}+ T_{comm,clusters},
\label{eq:Tmcp}
\end{equation}
where  $T_{comm,clusters}$ is the time for communication between hosts 
in different clusters. It is expressed as
\begin{equation}
T_{comm,clusters} = 72(2t_{comm,net}+t_{comm,grape})(q-1)/p,
\end{equation}
where $t_{comm,net}$ and $t_{comm,grape}$ are the time to send 1-byte
date through the network interface of host and host-GRAPE interface,
respectively. The constant factor of 72 is the length of data for one
particle in bytes. The next factor of 2 comes from the fact that each
node needs to both send and receive the data.  The factor $q-1$
appears since one node receives data from $q-1$ other nodes. We used
$t_{comm,net}=1.7\times 10^{-8}{\rm s}$ and $t_{comm,grape}=1.25\times
10^{-8}{\rm s}$. These values are based on separate measurement using
small benchmark programs. Figure \ref{fig:np16theory} shows the
calculation time per one particle step for full-cluster calculation
(16 nodes, 4 clusters).  The agreement between the theoretical
estimate and the measured value is fairly good, but not ideal. We
probably have underestimated $t_{comm,net}$ in the real program.

\subsubsection{Summary for the direct summation code}

In this section we presented the performance figures of GRAPE-6 for
the direct $N$-body simulation. As described in the introduction, this
kind of simulations is the main target of GRAPE-6. So we made fairly
detailed analysis of the performance.  What we have found is
summarized as follows.

In the case of the single-host configuration, the calculation speed of
the present host computer is the largest bottleneck of the
performance, and the communication speed is relatively
unimportant. This means that we can keep improving the overall
performance of the system just by replacing the host computer, for the
next several years.

For the multi-host configuration, the situation is rather
different. In the case of a single cluster (no host-host data
transfer), the performance for small-$N$ runs is determined by the
overhead of the barrier synchronization between host computers. We
currently use standard UNIX implementation of TCP/IP socket for the
basic communication, and TCP/IP socket is certainly not the
communication software with lowest possible latency. The use of
communication software/hardware with lower latency would significantly
improve the performance.

Finally, for the case of multi-cluster configuration, as expected, the
performance is limited by the  bandwidth of the communication between
hosts. Currently, we use Gigabit Ethernet card on 32-bit, 33-MHz
PCI bus. Clearly, by going to faster bus (PCI-66 or PCI-X) and faster
CPU, the communication bandwidth will be improved significantly.

To summarize, at present the performance of GRAPE-6 for small-$N$
calculations is limited by the speed of the host computer, and by the
latency of the communication between hosts when multi-host or
multi-cluster systems are used. Even so, GRAPE-6 can achieve the speed
exceeding 100 Gflops, for relatively small number of particles such as
16k. In the coming several years, the improvement of the host
computer will improve the overall performance of the system.

\subsection{Tree algorithm}

Here we discuss the performance of GRAPE-6 for Barnes-Hut tree
algorithm. We used the modified algorithm introduced by 
\citet{Barnes1990}. We discuss the performance of single-host code
and multi-host (parallel)  code with up to 12 host computers.  The parallel
algorithm is based on the space decomposition similar to the
well-known orthogonal recursive bisection (ORB) method
\citep{Dubinski1996}.  The detail of the parallel algorithm will be
discussed elsewhere.

\subsubsection{single-host performance}

\begin{figure}
\begin{center}
\leavevmode
\FigureFile(6 cm,6 cm){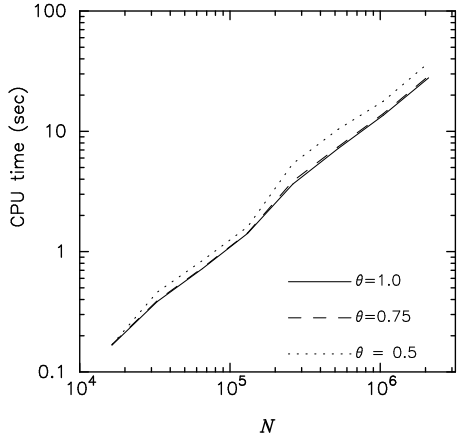}
\end{center}
\caption{CPU time in seconds per timestep plotted as a function of the
number of particles $N$ on a single-host configuration. Solid,
dashed and dotted curves are for $\theta=1$, 0.75, 0.5, respectively.}
\label{fig:treespeed1}
\end{figure}

Figure \ref{fig:treespeed1} shows the CPU time per timestep as a
function of the number of particle $N$. The distribution of particles
is a Plummer model, with the outer cutoff radius of 22.8 in Heggie
units. We used $n_g = 20,000$ as the maximum group size for the
modified algorithm.

We can see that the CPU time grows practically linearly as we increase
$N$. Also, the dependence on the opening angle $\theta$ is rather
weak. This weak dependence is the characteristic of GRAPE
implementation of the tree algorithm \citep{Makino1991c,Athanassoulaetal1998}.

Table \ref{tab:treetime} gives the breakdown of the CPU time per step
for calculation with $N=2^{21}$.  The average length of the
interaction list was $1.01\times 10^4 $ and $1.69\times 10^4 $ for
$\theta=1.0$ and $0.5$, respectively. The number of groups is $310$
for both cases. As in the case of tree algorithm on older GRAPE hardwares,
the performance is limited  by the speed of the host and that of
communication.  Actual calculation on GRAPE-6 takes less than
three seconds, for the case of $\theta = 0.5$. The calculation on the
host (tree construction, tree traversal, and other calculations
including the data conversion between GRAPE-6 internal format and
floating-point format) count for roughly 2/3 of the remaining time,
and actual communication 1/3. This, again, implies there is a rather
large room for the improvement of the speed, just by moving to faster
host computers.

\begin{table}
  \caption{CPU time distribution for treecode}\label{tab:treetime}
  \begin{center}
    \begin{tabular}{lcc}
\hline
Operation & Time (sec, $\theta=1$) &  ( $\theta=0.5$)\\
\hline
Tree construction & 7.57  & 7.57 \\
Force calculation & 18.40 & 27.62 \\
Other operations  & 1.86  & 1.86  \\
Total             & 27.83 & 37.05\\
\hline
    \end{tabular}
  \end{center}
\end{table}

\subsubsection{multi-host performance}

Since the performance of the single-host GRAPE-6 is limited by the
speed of the host computer, an obvious way to improve the performance
is to use multi-host systems.

\begin{figure}
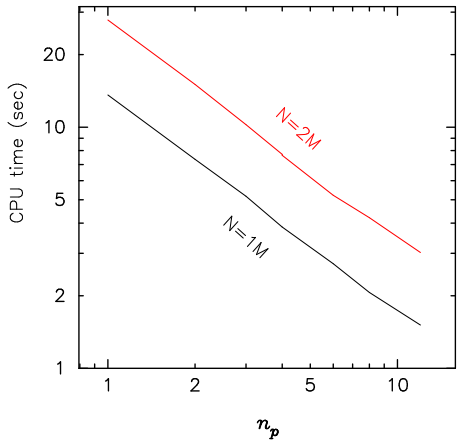

\begin{center}
\leavevmode
\FigureFile(6 cm,6 cm){ptreespeed2.ps}
\end{center}
\caption{CPU time in seconds per timestep plotted as function of the
number of host computers $n_p$. Parallel tree algorithm is used with
$\theta=1.0$. Upper and lower curves are the result for $2^{21}$ and
$2^{20}$ particles, respectively. Initial distribution of the
particles is a Plummer model.}
\label{fig:ptreespeed1}
\end{figure}

Figure \ref{fig:ptreespeed1} shows the performance of the parallel
tree algorithm. The program used is a newly written one based on orthogonal
recursive multi-section, a generalization of widely used ORB tree
that allows a division to arbitrary number of domains in one
dimension, instead of allowing only bisection. The primary advantage
of this algorithm is that it can be used on systems with number of
host computers not exactly a power of two. We measured the performance on
1, 2, 3, 4, 6, 8 and 12 hosts. 

The distribution of the particles is again the Plummer model. One can
see that the scaling is again pretty good. 12-hosts calculation is
9.3 times faster than single-host calculation. Parallel efficiency is
better than 75\%, even for relatively small number of particles shown
here.

%
% 20030329
%

\section{Discussion}

\subsection{Hindsights}

Though we regard GRAPE-6 a reasonable success, this certainly does not
mean we did everything right. We did make quite a
few mistakes, some of them affected the performance, some affected the
reliability, some extended
the development time,  and some limited the
application range. In the following we briefly discuss them in turn.

\subsubsection{Performance}

Concerning the performance, the largest problem with GRAPE-6 is that
its clock frequency is somewhat below the expected value. The design
goal (for the ``worst case'') was 100 MHz, while our actual hardware is
currently running at 90 MHz. With GRAPE-4, the design goal was 33 MHz,
and the machine operated without any problem at 32 MHz. The processor
chip itself was confirmed to operate fine at 41 MHz.

The primary reason for the low operating frequency is the problem with
the stability of the power supply to the chip, or the impedance of the
power line. Compared to the GRAPE-4 processor chip, GRAPE-6 processor
chip consumes about two times larger power, at half the supply
voltage. Thus, to keep the relative drop of the supply voltage to be
the same, the impedance of the power line must be 1/8 of that of
GRAPE-4.

This is a quite difficult goal to achieve. Initially even the
manufacturer of the chip did not fully appreciated how hard it was. As
a result, the first sample of the chip could not operate correctly at the
clock speed higher than 60 MHz. The problem was that, when calculation
starts, the power consumption of the chip increases by about a factor
of two compared to that when the chip is idle. Because of the large
total resistance of the power line in the LSI package and the power plane of
the silicon chip itself, the supply voltage to the transistors
decreases, and as a result their switching speed slows down.

In the second design, the manufacturer came up with additional power
plane and increased number of power and ground pins, which 
reduced the resistance significantly. However, the result was still
rather unsatisfactory.

The manufacturer was not alone in making this kind of mistakes. In the
design of the processor board and the power supply, we also made
similar mistakes. In the first design, we used traditional large
switching power units with relatively low switching frequency. This
unit turned out to be unable to react to the quick change of the load
between idle and calculation states. The normal electrolytic capacitors
also turned out to be completely useless in stabilizing the power
supply voltage. Thus, we need to redesign the power supply unit with
high-frequency inverters and low-ESR capacitors. 

In hindsights, we could have borrowed the design of power supply units
for standard PC motherboards (for Intel processors), which were
designed to meet quite similar requirements, but for an extremely low
cost. The power supply circuit for typical PC motherboard would be
good enough to support single module with 4 chips. We then can supply
12 V to PCB.

These apparently minor technical details are absolutely crucial for
the manufacturing of high-performance computers.

Another problem with the current GRAPE-6 chip is its limited I/O
performance of only 90MB/s. As we stated earlier, this bandwidth is
sufficient to keep the standard PCI interface busy, and it is not
really the bottleneck, since, for many applications, the calculation
on the host computer is more time-consuming. Even so, in a few years
the I/O performance will become a problem. Additional problem is that
host computers with faster PCI interfaces (PCI64 and PCI-X) are now
available. We cannot take advantage of these faster interfaces with
the current GRAPE-6 design, because the I/O bandwidth of the processor
chip is limited. We could have increased the I/O bandwidth of the chip
without too much problem, by allowing the change in the ratio between
the chip clock and board clock. With our current design, this ratio is
effectively fixed to 4.

Even with the current chip design, we could have increased the
communication bandwidth of the processor board, without increasing
that of the chip, by letting multiple chips to transfer the data
simultaneously. This possibility should have been considered, to
increase the lifetime of the hardware.

\subsubsection{Reliability}

Since the GRAPE-6 system consists of exceptionally large number of
arithmetic units, one might imagine that the primary source of the
error is the calculation logic itself. In practice, however, we have
almost never seen any calculation error, once the power supply had
become good enough. On the other hand, we saw quite a few errors in
data transfer.

We implemented ECC circuit for the memory interface of the processor
chip, but only added parity detection circuit to I/O ports. We thought this is
reasonable, since memory ports operate on 90 MHz clock and I/O ports
on 22.5MHz. However, it turned out that memory parity error almost
never occur, while parity error for I/O occurs rather
frequently. Since we do not exactly know the type of the error, it is
not 100\% clear whether the ECC capability would have helped or
not. However, it is at least clear that more reliable data
transmission would be better.

A more serious problem with the reliability was very high defect rate
for mass-produced processor boards and processor modules. Practically
all failures were due to unreliable soldering, and most of soldering
problem turned out to be simply due to lack of skill of the
manufacturer. This may be telling something about the present performance
of Japanese high-tech industry. Even so, it is certainly true that to
manufacture a rather small quantity of PCB board is difficult. 
We could have designed either
more automated test procedures for boards (with JTAG standard) or
redundant connection. Yet another possibility was to reduce the
number of wires by using higher-frequency signals.

\subsubsection{Development time}

As we discussed in section 5, the use of the parallel host was
inevitable. However, the use of the multicast network was not, at
least in hindsight. We assumed that the price of high-end uniprocessor
computers would not change greatly, and that the cost of
high-bandwidth network adapters (1 Gb/s or higher) would remain high. In
other words, we assumed that we could not afford to buy $\sim$ 100
fast host computers and to connect all of them by a fast
network. Therefore, we designed our own network, which connects $r$
host computers to $r^2$ processor boards. This approach worked fine,
as we have seen in the previous section. However, an alternative
design, in which we connect each processor board to its own host
computer, would have been much easier to develop.

In 1997, the fastest systems are RISC-based UNIX workstations, with
price higher than 20K USD. In 2003, systems based on the Intel x86
architecture offer the speed similar to that of the RISC based systems
with highest performance, for the cost of less than 2K USD. TO
illustrate this, we
use SPECfp (either 95 or 2000) numbers as representative of the
performance. In 1997 the speed difference between RISC systems and x86
systems were nearly a factor of three. This ratio had been almost
constant during 1990s. The reason why the ratio started to shrink is
simply that the rate of improvement of the performance of RISC-based
systems slowed down. Thus, it would have been difficult to predict the
present state in 1997, or even in 1999. In other words, even though
now it is clear that network hardware of GRAPE-6 is not necessary,
until 2000 we had no other choice.

Another reason for the rather long development time (aside from the
problems with the power supply) is the fact that we integrated
effectively all functions of the system to the processor chip. It
integrates the memory controller, the predictor pipeline, and all
other control logics. Except for the predictor pipeline, all these
logics were implemented with FPGAs on GRAPE-4. This integration 
simplified the design of the board, and fortunately, we have not made
any serious mistake in the design of these parts. The integration of
these complicated logics onto hardware required extremely careful
design and test procedures which were time-consuming. With the present
price of moderately large FPGA chips, all of these control logics
could be implemented using FPGAs, with very small additional cost. Of
course, such moderately large and inexpensive FPGA was not there when
we decided on the design of GRAPE-6 chip. However, we could have
predicted the direction of the evolution of FPGA chips and estimated
the price of them.

\subsubsection{Application range}

Since GRAPE-6 is designed solely for the gravitational $N$-body
problem, one might think there is not much of the range of
applications. However, even within $N$-body simulations, there are
many factors.

The overall design of GRAPE-6 is highly optimized to parallel
execution of the direct force calculation with the individual timestep
algorithm. This of course means it is not optimal for other
applications, such as tree algorithm and SPH calculations of
self-gravitating fluid.

With the case of the tree algorithm, the performance is limited mainly
by the speed of the host computer. So, in this case, adding more host
computers would have greatly improved the performance. 

In principle we could have improved the performance of the tree
algorithm in several other ways. One obvious approach is to reduce the
data to send. With tree algorithm, we would not use the
predictor. Moreover, we would not need the full 64-bit resolution for
the position data. Thus, we could have implemented some way to reduce
the data to send for $j$-particles, if our memory controller was not
implemented in hardware. Actually, the memory controller of GRAPE-6
has some programmability. However, one ``feature'' of this memory
controller prevented us from taking full advantage of this
programmability to reduce the amount of the data transfer.

With an FPGA implementation of the memory controller, we could implement
other ways to further reduce the communication. For example, we could
implement indirect addressing, so that we can send indices of $j$
particles instead of sending their physical data.

Concerning the design of the pipeline, one thing which might have been
useful for simulation of collisionless systems or composite
$N$-body+SPH systems is the ability to apply different softening
length on different particles, in a symmetrized way. This can be
achieved by calculating the softened distance as
\begin{equation}
r_s^2 = r_{ij}^2 + \epsilon_i^2 +\epsilon_j^2.
\end{equation}
The pipeline will need one more addition, which is relatively
inexpensive. 

With SPH, the main problem is that the calculation of SPH interactions
itself cannot be done on GRAPE-6. The PROGRAPE system\citep{Hamadaetal2000}, with
the calculation pipeline fully implemented in FPGA, could be used to
perform the calculation of SPH interaction. Moderately large PROGRAPE
system is currently under development.

With the logic design of the pipeline, we have noticed a few problems
which we could not foresee. One is the length of the accumulator for
the time derivative of the force. For the force and potential, we used
64-bit accumulators, but for the time derivative we used 32-bit
accumulators. As far as the accuracy is concerned, this length is long
enough. However, when we performed simulations with large number of
particles, we realized that the overflow occurred rather
frequently. The reason why the overflows occurs is that the magnitude
of the time derivative of the force can change by a large factor in a
single timestep. The large change occurs when the previous value
happens to be almost zero. We could circumvent this problem with
a combination of guess for the likely value of the time derivative of
the force based on the value of force and timestep, but it is
cumbersome to implement and expensive to evaluate. By increasing the
accumulator length to, say, 40 bits, we could have almost completely
eliminated the overflow.  This overflow does not have any noticeable
impact on performance. But the need to handle overflows made the
interface program rather complicated.

\subsection{GRAPE-7/8}

Given that GRAPE-6 is now completed and we already have the experience
of running it for almost two years, it would be natural to put some
thought on how its successor will look like. In this section, we first
discuss the change in the technologies, and then overview the design
possibilities.

\subsubsection{Technological changes and the basic design}

Compared to the technology used in GRAPE-6, what will used in the next
GRAPE system (we call it NGS for short) will be different in

\begin{description}
\item (a) Semiconductor technology 
\item (b) development cost
\item (c) Host I/O bus
\end{description}

First let us discuss the semiconductor technology. GRAPE-6 used
$0.25{\rm \mu m}$ technology, while NGS would use, depending on the
time to start, either $130 {\rm nm}$ or $90 {\rm nm}$
technology. Since it seems we are not getting the budget too soon, we
will probably use $90 {\rm nm}$. This means we can pack about 8 times
more transistors to the chip of the same size, and the switching speed
will be about 3 times faster. Thus, a single chip of the same size can
offer 20 times more computing power.  If the power supply voltage is
reduced by the same factor, the power consumption would remain the
same, but most likely the supply voltage would be somewhat higher,
resulting in significant increase in the power consumption.

To express in concrete numbers, a single chip would integrate around
50 pipeline processors, each with 60 arithmetic units operating on
300 MHz clock speed, with 1.2V supply voltage and power consumption of
20 W. The theoretical peak speed of the chip will be around 600
Gflops.

Compared to the projected speed of general-purpose microprocessors in,
say, 2007, this speed is quite attractive. In 2007, microprocessors
will, at best,  have the peak speed 10 times faster than they have
now, or about 30 Gflops. Typical performance on real application would
be around 10 Gflops or less, for the power consumption of 100 W or
more.

A necessary consideration is how we connect the pipelines to
memory. If we use the same memory system as we used for GRAPE-6, the
total number of virtual pipelines per chip becomes 1,000, which is 
too large for a simulation of any collisional system. As was the case
with GRAPE-6, it is necessary to keep the number of $i$-particles
calculated in parallel to be around 500 or less, for large systems with
many chips. So the number of virtual pipelines per chip must be less
than 200, or ideally less than 100. In other words, the memory bandwidth
must be increased by at least a factor of five, to around 3.5 GB/s.

This number, by itself, sounds relatively easy to achieve. It is the
same as what was used with the first Intel P4 processor (3.2GB/s),
using two DRDRAM channels each with 16-bit data width. Intel P4 has
been around for more than two years. Now we can also use DDR 400
memory chips, which have 4 times more throughput than the SSRAM chips
used in GRAPE-6. We could also use DDR SRAMs.

The choice of the memory interface has strong impact on the range of
the applications. One major limitation of GRAPE-6 was that, as was
discussed in the previous section, its memory addressing scheme was
limited only to the sequential access to a full set of predictor
data. Thus, it is not easy to use the tree  or other
sophisticated algorithms efficiently on GRAPE-6. One possibility to
solve this problem is to implement the memory controller and other control
logics in an FPGA chip. The connection between the FPGA chip and the
pipeline chip must be quite fast, but this is relatively easy to
achieve since the data transfer is unidirectional, from the FPGA chip
to the pipeline chip.  The memory controller will be implemented in
the FPGA chip.  Thus, it will be possible to use different types of
memory (DRDRAM, DDR DRAM/SRAM) without any need to change the pipeline
chip.

As we discussed earlier, parallelism will be achieved by
two-dimensional network of host computers. Each of them will have a
relatively small GRAPE system. As an example, we consider a system
with 256 host computers each with two GRAPE cards. Each card houses 4
processor chips with their own memory control units and memories. All
of them can be packaged into single card of the PCI form factor, though we
need special care for the power supply.

For the interface to the host, the easiest solution is to use PCI-X,
which is available now with the data transfer speed of up to 1
GB/s. PCI-X gives us an order-of-magnitude increase in the
communication speed, which roughly balances the increase in the
performance of a factor of 20.  One problem is whether or not PCI-X
will be around 5 years from now. We need to predict the market trend,
or develop a design that can use multiple interfaces.

Note that this factor-of-10 increase in the communication implies that
the chip-to-chip communication must also be faster by the same
factor. This is not easy, but since the physical size of the board
will be much smaller, it would not be impossible to use fast clocks.

Thus, the design of NGS seems to be  simple, as far as we
set the parallel execution of the individual timestep algorithm with
direct summation as the primary design target.

The only, but quite serious, problem is that the predicted initial
cost for the custom chip will be very high. The initial cost for a custom
chip has been increasing quite steeply. Roughly speaking, the initial
cost has been proportional to the inverse of the design rule. Thus,
while the initial cost of the GRAPE-4 chip was around 200K USD, that
for GRAPE-6 exceeded 1M, and for NGS it will reach 3M. Even though
this is ``small'' compared to the price of any massively-parallel
supercomputer or even PC clusters, to get a grant of this size
within the small community of theoretical astrophysics in Japan is not
easy.

\subsubsection{Combination with sophisticated algorithms}

One rather fundamental question concerning the next GRAPE system is
whether the direct summation is really the best solution or
not. McMillan and Aarseth (\yearcite{McMillanAarseth1993}) have
demonstrated that it is possible to implement a combination of the
Barnes-Hut tree algorithm and the individual timestep algorithm that
runs efficiently at least on a single-processor computer, and
potentially also on shared-memory parallel computers. Even when we
require very high accuracy, the gain by tree algorithm is large for
large $N$. For example, the number of interactions per particle to achieve
the relative force accuracy of $10^{-5}$ is around 8,000 when
quadrupole moment is used and around 2,000 when octupole moment is
used, for number of particles around $10^6$. Thus, even if we assume
the calculation of octupole costs a factor of 10 more than point-mass
force, the calculation cost of the tree algorithm would be a factor of 50
less than that of the direct calculation.

Even though the scaling is not as drastic as that of the tree
algorithm, the Ahmad-Cohen scheme (\yearcite{AhmadCohen1973}, also known
as the neighbor scheme) offers quite significant reduction of the
calculation cost over the simple direct summation. The theoretical
gain in the calculation cost is $O(N^{1/4})$ for the neighbor
scheme\citep{MakinoHut1988,MakinoAarseth1992}. However, the actual
speedup is nearly a factor of 10, for only 1,000 particles. thus, for
$10^6$ particles the gain can reach a factor of 50.

For $10^6$ particles both the tree algorithm and neighbor
scheme, at least theoretically, offer the reduction in the calculation
cost of around a factor of 50. This factor is certainly still smaller
than the advantage of the GRAPE hardware over general-purpose
computers, since the difference in the price-performance ratio will
exceed $10^3$. However, if we can incorporate either of these
sophisticated algorithms, even with significant loss in the hardware
efficiency like a factor of 5 or even more, we can still achieve a very
significant improvement in the overall speed. We are currently
investigating several possible ways to achieve this goal.

\section*{Acknowledgments}

We would like to thank all of those who involved in the GRAPE
project. In particular, We thank Daiichiro Sugimoto for his continuous
support to the project,  Atsushi Kawai for helping the
hardware design, Yoko Funato, Eiichiro Kokubo, Simon Portegies Zwart,
Piet Hut, Steve McMillan, Makoto Taiji, Sverre Aarseth, Takayuki Saito and many others for
discussions on the experience with GRAPE-4 and 5. We are grateful to Mary
Inaba and Junichiro Shitami for discussions on the network
performance, and for providing us their software to measure the
network performance.   This work is supported by
the Research for the Future Program of Japan Society for the Promotion
of Science (JSPS-RFTF97P01102).

%\bibliographystyle{../bibtex/apj} %*** Key this instruction into your document
%\bibliography{../bibtex/allrefs}    %*** Supply the name of your
%bibliography file.
%\end{document}

\newcommand{\noopsort}[1]{} \newcommand{\printfirst}[2]{#1}
  \newcommand{\singleletter}[1]{#1} \newcommand{\switchargs}[2]{#2#1}

\end{document}